\documentclass[
aps,prmaterials,reprint,superscriptaddress,
longbibliography,floatfix
]{revtex4-2}

\usepackage{amsmath,amssymb,bm}
\usepackage{graphicx}
\usepackage{dcolumn}
\usepackage{xcolor}
\usepackage{microtype}
\usepackage{hyperref}
\usepackage[normalem]{ulem}  
\hypersetup{hidelinks}

\begin{document}

\title{Unsupervised autoencoder resolves crystal like order and structure dynamics correlations in glass forming Kob Andersen mixtures across compositions}


\author{Sanket Kumawat}
\affiliation{Polymer Science and Engineering Division,
CSIR-National Chemical Laboratory, Pune 411008, India}
\affiliation{Academy of Scientific and Innovative Research (AcSIR), Ghaziabad 201002, India}

\author{Dibyendu Maity}
\affiliation{Department of Chemical and Biological
Sciences, S. N. Bose National Centre for Basic Sciences,
Kolkata 700106, India}

\author{Mohit Sharma}
\affiliation{Polymer Science and Engineering Division,
CSIR-National Chemical Laboratory
, Pune 411008, India}
\affiliation{Academy of Scientific and Innovative Research (AcSIR), Ghaziabad 201002, India}

\author{Krishna Rijal}
\affiliation{Department of Organismic and Evolutionary Biology,
Harvard University, Cambridge, MA 02138, USA}

\author{Ujjwal Kumar Nandi}
\affiliation{Department of Physics, Kyoto University,
Kyoto 606-8502, Japan}
\affiliation{Research and Business Development Office, Tokyo Electric Power Services Co. Ltd, Tokyo, Japan}

\author{Suman Chakrabarty}
\affiliation{Department of Chemical and Biological
Sciences, S. N. Bose National Centre for Basic Sciences,
Kolkata 700106, India}

\author{Sarika Maitra Bhattacharyya}
\email{Corresponding author: mb.sarika.ncl@csir.res.in}
\affiliation{Polymer Science and Engineering Division,
CSIR-National Chemical Laboratory
, Pune 411008, India}
\affiliation{Academy of Scientific and Innovative Research (AcSIR), Ghaziabad 201002, India}


\begin{abstract}
    Understanding how local structure encodes slow dynamics in supercooled liquids, and how it relates to competing crystalline order, remains a central challenge. We study Kob Andersen binary Lennard Jones mixtures with B particle fractions $\chi_B=0$ to $0.5$, whose frustration towards crystallisation varies with composition. Averaged smooth overlap of atomic positions (SOAP) descriptors are compressed into a two dimensional latent space by autoencoders trained without dynamical information or crystal labels. When cooling trajectories are projected onto this space, the Residual autoencoder places the crystallising compositions in the fcc/hcp and CsCl basins and keeps the frustrated compositions $\chi_B=0.2$-$0.4$ in the liquid region. Unlike bond orientational order, it also separates liquid states at different temperatures and compositions, which we attribute mainly to the radial information in SOAP. The learned coordinates correlate with isoconfigurational mobility in independently simulated liquids, whereas the bond orientational order does not. Among the architectures tested, the Residual autoencoder and the Graph autoencoder give the strongest correlations. The learned coordinate closely tracks the microscopic mean field caging potential, with which it correlates at $0.49$ for bare values and $0.89$ after coarse graining. Spatial coarse graining of the machine learned order parameter further enhances its correlation with dynamics at low temperatures, with a maximum at a finite length scale that increases on cooling.
\end{abstract}

\maketitle

\section{Introduction}
\label{sec:introduction}

When a liquid is cooled fast enough, it bypasses crystallisation, enters the supercooled regime, and, on further cooling, becomes a glass \cite{Cavagna2009PhysRep,Angell1995Science,
Debenedetti1996Book,DebenedettiStillinger2001Nature,BerthierBiroli2011RMP}. Besides fast cooling, other mechanisms can suppress crystallisation and promote supercooling. One of them is frustration, which can be geometric, entropic, or energetic in nature. A classic example is due to Frank \cite{Frank1952}, who proposed that icosahedral order is the locally preferred structure of simple liquids and that it is incompatible with crystalline order. The liquid is therefore locally stable in an icosahedral arrangement, whereas the crystal is globally stable. The costly rearrangement between the two structures slows the crystallisation and allows the liquid to be supercooled. This mismatch between local order and the tiling of space is known as geometric frustration \cite{Tarjus2005}.

In the Kob Andersen (KA) binary mixture, the strength of frustration depends on composition. A-rich mixtures tend to form fcc order, whereas the equimolar mixture forms CsCl like order \cite{atreyee_frust,Ujjwal_frust,Pedersen2018PRL}. Between these limits, the phase diagram is V-shaped, and, as in other liquids with a V-shaped phase diagram \cite{sastry_molinero,tanakavshape,Tanaka2012EPJE}, the crystallisation barrier is highest near the bottom of the V, $A:B=80:20$ ($\chi_{B}=0.2$). In this system, one of the lowest energy crystal structures is a coexistence of CsCl (formed by A and B particles) and fcc (formed by the remaining A particles) \cite{FernandezHarrowell2003PRE}, and the demixing entropy required to form it makes crystallisation difficult \cite{Ujjwal_frust}. This is why the standard $\chi_{B}=0.2$ KA mixture is a good glass former, although it has a tendency to form fcc order \cite{Ujjwal_frust}.  In long simulations, A particles can crystallise into fcc if a sufficiently A-rich region forms \cite{jeppe_frust,Pedersen2018PRL}.


The degree of frustration has been suggested to play an important role in the dynamics of supercooled liquids, and to control how static and dynamic length scales are related \cite{Charbonneau2012PRL,Tah2018PRL}. In weakly frustrated liquids with medium range crystalline order, the two lengths grow together, whereas in strongly frustrated liquids they decouple. For the $\chi_{B}=0.2$ mixture, no growing medium range crystalline order was observed, and the static lengths identified so far do not track the dynamical one. For example, the networks of long lived eleven particle clusters identified using topological cluster analysis (TCC) by Royall and co-workers grow more slowly on cooling than the dynamical correlation length \cite{Malins2013Faraday,Flenner2014PRL,Berthier2016JCP}.
The order agnostic point to set correlation length for this system also did not grow enough \cite{Berthier2016JCP,Yaida2016PRE}.
These observations suggest that structure may play a weaker role in the dynamics of more frustrated systems than of less frustrated ones, and the standard KA model belongs to the former group. 

In a series of work involving some of us, \cite{manoj_PRL2021,Sharma2022Softness,
Sahu2024PNAS} it was shown that the local structure of the liquid can be described by a local mean field caging potential obtained from microscopic density functional theory \cite{Schweizer2005JCP}. This caging potential gives a good structure dynamics correlation in different systems \cite{manoj_PRL2021,mohit_LJ_WCA, Patel2023JCP,Patel2024JCP,Anwar2024JCP,Sahu2024PNAS}, including the standard KA model \cite{Sharma2022Softness,Kumawat2025MRO}. This indicates that even in the standard $\chi_{B}=0.2$ KA model, the structure has a strong influence on dynamics, provided the structural information is captured by a suitable order parameter and coarse grained over a suitable length.

Machine learning has also been used extensively to relate structure to
dynamics in the KA liquid both in 3D and 2D. In the 2D KA liquid, the ML order parameter predicts a very good structure dynamics correlation, even with an unsupervised learning method \cite{SharmaOzawa2024JCP}. Here, we primarily discuss the results obtained in the 3D systems.  Many successful studies have used supervised approaches, in which local structure is fitted to particle dynamics that must be supplied during training. Amongst them, Liu and co-workers  ~\cite{andrea_NatPhys2016} used the support vector machine and showed that
the probability of rearrangement increases with the softness, a
structural variable learned from the local particle environment.
Particles with more structured neighbours tended to have lower
softness, and lower mobility, consistent with the cage picture \cite{manoj_PRL2021,Sharma2022Softness,
Sahu2024PNAS,Kumawat2025MRO}.
 Bapst \textit{et al.}~\cite{Bapst2020NatPhys} predicted the
dynamic propensity of particles 
using a graph neural network that passes information between
neighbouring particles.
Boattini \textit{et al.}~\cite{Boattini2021PRL} achieved accuracy
comparable to that of the graph neural network by using linear
regression on local structural descriptors that were recursively
averaged over neighbouring particles. 
Alkemade \textit{et al.}~\cite{Alkemade2023Cage} used linear
regression to predict dynamic propensity and found that structural
descriptors of the cage state were particularly informative
at long times. Berthier and co-workers~\cite{Jung2023PRL} used a supervised
neural network to predict particle mobility from physics inspired
structural descriptors coarse grained over several length scales.
In a subsequent study~\cite{JungBiroliBerthier2024PRB}, they
examined the relative importance of these length scales and found
that intermediate range structural information becomes
increasingly important upon cooling. Coslovich {\it et al.} used a linear regression method and compared different structural input parameters (smooth bond order (SBO), smooth overlap of atomic positions (SOAP)) and found that SOAP performs better than SBO.
Shiba \textit{et al.}~\cite{Shiba2023JCP} trained a graph neural
network using propensity and the distances between neighbouring particles to predict structure dynamics correlation. Please refer to Ref.~\cite{Jung2025Roadmap} for a detailed comparison of these approaches.

Unsupervised approaches, in contrast, identify structural patterns without using dynamical information. Any subsequent correlation with particle mobility is therefore an independent finding.
Paret \textit{et al.} used an information theoretic approach to group particles having similar two and three body static correlations and identified local orders \cite{Paret2020JCP}. In their study, the coarse grained local density performed the best in describing the dynamics. Boattini \textit{et al.} used an
autoencoder where the input is bond orientational order parameters and showed that their order parameter has good correlation with the dynamics \cite{Boattini2020NatCommun}.
Coslovich \textit{et al.} showed that principal component analysis
(PCA) and autoencoders recover equivalent information from
bond orientational descriptors ~\cite{Coslovich2022JCP}. The SBO descriptor did not perform well in predicting the dynamics; however, the coarse grained SBO descriptor performed much better \cite{Coslovich2022JCP,Jung2025Roadmap}.
Madanchi and Simine trained a generative model on static data and showed that a score built from its conditional position probabilities tracks dynamic excitations \cite{Madanchi2026MAP}. Yethiraj \textit{et al.} used a time lagged autoencoder (TAE) where they trained on radial and orientational features of configurations 
separated by a short time lag and found a structural coordinate that
correlates very well with mobility on the scale of the relaxation time, with
density variations in the third coordination shell and beyond being
especially informative~\cite{QiuYethiraj2025PNAS}. Amongst the unsupervised methods, Yethiraj {\it et al.} study shows the best structure dynamics correlation. 

In this work, our first goal is to develop a structural order parameter to study the crystallisation tendencies of the family of KA models by varying their composition, which in turn varies the degree of frustration towards crystallisation. We then want to study the prediction of this structural order parameter in the dynamics of the systems which are strongly frustrated towards crystallisation. As discussed before, the microscopic caging potential is a useful structural order parameter and correlates well with the dynamics. The caging potential can also distinguish a crystal from a liquid \cite{Sahu2024PNAS}. However, it cannot be directly connected to crystallisation tendencies in the system. We show that the Bond orientational order (BOO) parameters, which predict crystallisation tendencies, do not resolve the precursors of crystalline order when crystallisation is frustrated and do not predict dynamics well. Motivated by a recent study where different phases of crystalline ice and liquid water were identified using a variational autoencoder \cite{MaityChakrabarty2025JCTC},
we use various autoencoder (AE) models to identify crystal and pre crystalline motifs in different systems. We train the models by feeding averaged SOAP vectors obtained for different temperatures and different compositions. We show that the models thus trained can predict transitions in systems with different compositions that undergo crystallisation, and clearly reveal a temperature and composition variation of the order parameter in systems that are frustrated towards crystallisation. Thus, these AE models are quite generic and work across temperatures and compositions. Among the various models, the residual autoencoder appears to be the simplest with the best predictive power. We find that the order parameter from the residual AE model correlates well with the microscopic caging potential derived from theory \cite{manoj_PRL2021,Sharma2022Softness}. It also successfully shows a good structure dynamics correlation, which further improves with the coarse grained order parameter.

The paper is organized as follows. Section~\ref{sec:model} describes the Kob Andersen model and simulation protocol. Section~\ref{sec:dynamics} defines the isoconfigurational mobility and relaxation time, and Section~\ref{sec:structure} defines the structural descriptors, the autoencoder architectures, and the microscopic caging potential. The results are discussed in Section~\ref{sec:results}. Section~\ref{sec:comparison} compares our approach with existing unsupervised studies, and Section~\ref{sec:conclusions} concludes. Training details and additional results are given in the Supplemental Material (SI).

\section{Model and simulation protocol}
\label{sec:model}
\subsection{Kob Andersen binary Lennard Jones model}

We study the three dimensional Kob Andersen (KA) binary
Lennard Jones mixture~\cite{KobAndersen1994PRL,KobAndersen1995PRE},
consisting of particles of species A and B. For a pair of
particles of species $\alpha$ and $\beta$, the Lennard Jones
potential before truncation and shifting is
\begin{equation}
U_{\alpha\beta}^{\mathrm{}}(r)
=
4\epsilon_{\alpha\beta}
\left[
\left(\frac{\sigma_{\alpha\beta}}{r}\right)^{12}
-
\left(\frac{\sigma_{\alpha\beta}}{r}\right)^6
\right],
\end{equation}
where $r$ is the distance between the two particles and
$\alpha,\beta\in\{\mathrm{A},\mathrm{B}\}$.
The interaction parameters are
$\epsilon_{AA}=1.0$, $\epsilon_{AB}=1.5$,
$\epsilon_{BB}=0.5$, $\sigma_{AA}=1.0$,
$\sigma_{AB}=0.8$, and $\sigma_{BB}=0.88$.
For all pair types, the potential is truncated at
$r_{\rm c}^{\alpha\beta}=2.5\sigma_{\alpha\beta}$ and shifted to zero at the cutoff.
All quantities are expressed in reduced Lennard Jones units.
The composition is specified by $\chi_B=N_B/N$, where $N_B$
is the number of B particles and $N$ is the total number
of particles.

For the composition dependent cooling studies, we perform
molecular dynamics simulations with $N=16\,000$ particles
in the isothermal isobaric (NPT) ensemble at pressure
$P=10.19$, using an integration time step of $\Delta t=0.005$.
The temperature is controlled using a Nos\'e-Hoover
thermostat~\cite{Nose1984JCP,Hoover1985PRA}.
Each system is initially equilibrated at $T=5.0$ and
subsequently cooled or prepared at selected target temperatures.

We also perform crystal liquid interface simulations in a
slab geometry, with a crystalline slab in contact with the
liquid. These simulations are used to examine crystal growth
and changes in the structure and composition of the adjacent
liquid.

For the structure dynamics analysis, we perform independent
simulations with $N=16000$ particles in the canonical (NVT)
ensemble at number densities $\rho=1.20$ and $\rho=1.41$
for the $\chi_B=0.2$ and $\chi_B=0.4$ mixtures, respectively.

\section{Dynamical representations}
\label{sec:dynamics}

\subsection{Isoconfigurational mobility and relaxation time}
\label{sec:mobility}

We use particle mobility to characterise the dynamics.
To reduce the dependence of the measured mobility on
the choice of initial velocities, we employ an
isoconfigurational ensemble. This consists of independent
trajectories initiated from identical particle positions
but with independently sampled initial velocities at
the same temperature. The mobility of a particle $i$
at time $t$ is defined as
\begin{equation}
\mu_i(t)
=
\frac{1}{N_{\mathrm{iso}}}
\sum_{k=1}^{N_{\mathrm{iso}}}
\left|
\mathbf r_i^{(k)}(t)-\mathbf r_i(0)
\right|^2,
\label{eq:mobility}
\end{equation}
where $N_{\mathrm{iso}}=100$ is the number of
isoconfigurational trajectories,
$\mathbf r_i^{(k)}(t)$ is the position of particle $i$
in trajectory $k$, and $\mathbf r_i(0)$ is its common
initial position. This averaging disentangles the
influence of the initial structure from fluctuations
associated with the initial velocities.

In this work, we primarily focus on the long time
mobility evaluated at the structural relaxation time,
$\tau_\alpha$, defined by $Q(\tau_\alpha)=1/e$,
where $Q(t)=\frac{1}{N}\sum_{i=1}^{N}
\omega(|\mathbf{r}_i(t)-\mathbf{r}_i(0)|)$
is the self overlap function.
Here, $\mathbf{r}_i(t)$ is the position of particle
$i$ at time $t$, and the overlap function
$\omega(x)$ takes the value 1 for $x\leq a$
and 0 otherwise. We choose a cutoff distance
$a=0.3\sigma$ to account for small particle
displacements arising from vibrations within
their local cages.

\section{Structural representations}
\label{sec:structure}
\subsection{Bond orientational order}

We compute bond orientational order (BOO) using the
spherical harmonic formalism of Steinhardt
\textit{et al.}~\cite{Steinhardt1983PRB}. For a particle
$i$ with neighbour set $\mathcal N_b(i)$ and coordination
number $N_b(i)$,
\begin{equation}
q_{\ell m}(i)
=
\frac{1}{N_b(i)}
\sum_{j\in\mathcal N_b(i)}
Y_{\ell m}(\hat{\mathbf r}_{ij}),
\end{equation}
where $\hat{\mathbf r}_{ij}$ is the unit vector from
particle $i$ to particle $j$. 

In this paper we use the neighbour averaged parameters introduced
by Lechner and Dellago~\cite{LechnerDellago2008JCP}:
\begin{equation}
Q_{\ell m}(i)
=
\frac{1}{1+N_b(i)}
\displaystyle\sum_{j\in\{i\}\cup\mathcal N_b(i)}
q_{\ell m}(j),
\end{equation}
and
\begin{equation}
 Q_\ell(i)
=
\left[
\frac{4\pi}{2\ell+1}
\sum_{m=-\ell}^{\ell}| Q_{\ell m}(i)|^2
\right]^{1/2}.
\end{equation}

We consider $\ell=4$ and $6$ to characterize
local crystalline ordering. Neighbour averaging
reduces the sensitivity of BOO parameters to
local fluctuations and improves the distinction
between crystalline and liquid environments.~\cite{LechnerDellago2008JCP}

\subsection{Topological cluster classification}
\label{sec:tcc}

We also classify the
local structure using the topological cluster
classification (TCC) algorithm developed by Royall and
co-workers~\cite{paddy2013TCC}. TCC identifies
clusters from the topology of the bond network formed by neighbouring
particles. Bonds are assigned using a modified Voronoi
construction, with the maximum bond length
chosen as the largest first minimum position
of the radial distribution function among
all state points studied.
The parameter $f_c$, which controls whether four membered rings are
distinguished from three membered rings, is set to $f_c = 1.0$,
which reduces the construction to the standard Voronoi definition of
direct neighbours. Particles whose bond networks are
topologically equivalent to a reference cluster are assigned to that
cluster. We track the FCC, HCP, 10B, and 13K motifs.
The population of each motif is represented by the fraction of particles belonging to the corresponding cluster out of the total particles, averaged over 100 equilibrium configurations at each temperature.

\subsection{Smooth Overlap of Atomic Positions}
\label{sec:soap}

We characterise local particle environments using Smooth
Overlap of Atomic Positions (SOAP)
descriptors~\cite{Bartok2013PRB}, as implemented in
\texttt{DScribe}~\cite{Himanen2020CPC}.
SOAP encodes radial and angular correlations through
an expansion of the local particle density in radial
basis functions and spherical
harmonics~\cite{Coslovich2022JCP}.

For a central particle $i$, the smooth local density
is defined as
\begin{equation}
\rho^{(i)}(\mathbf r)
=
\sum_{j\in\mathcal N(i)}
\exp\left[
-\frac{|\mathbf r-\mathbf r_{ij}|^2}
{2\sigma_{\rm SOAP}^2}
\right],
\label{eq:soap_density}
\end{equation}
where $\mathbf r_{ij}$ is the relative position of
particle $j$ with respect to particle $i$, accounting
for periodic boundary conditions.
The neighbourhood $\mathcal N(i)$ is defined by
the SOAP cutoff $r_{cut}$, and
$\sigma_{\rm SOAP}$ controls the spatial smoothing
of the density.

The density is expanded as
\begin{equation}
\rho^{(i)}(\mathbf r)
=
\sum_{n=1}^{n_{\max}}
\sum_{\ell=0}^{\ell_{\max}}
\sum_{m=-\ell}^{\ell}
c_{n\ell m}^{(i)}
g_{n\ell}(r)
Y_{\ell m}(\hat{\mathbf r}),
\label{eq:soap_expansion}
\end{equation}
where $g_{n\ell}(r)$ and
$Y_{\ell m}(\hat{\mathbf r})$ are orthonormal radial
basis functions and spherical harmonics, respectively.
The expansion coefficients are obtained by projection:
\begin{equation}
c_{n\ell m}^{(i)}
=
\int d^3\mathbf r\,
\rho^{(i)}(\mathbf r)
g_{n\ell}(r)
Y_{\ell m}(\hat{\mathbf r}),
\label{eq:soap_coefficients}
\end{equation}
Rotationally invariant power spectrum components
are then constructed as
\begin{equation}
p_{nn'\ell}^{(i)}
=
\sqrt{\frac{8\pi^2}{2\ell+1}}
\sum_{m=-\ell}^{\ell}
c_{n\ell m}^{(i)*}
c_{n'\ell m}^{(i)}.
\label{eq:soap_power_spectrum}
\end{equation}
The indices $n$ and $n'$ describe radial correlations,
while $\ell$ specifies the angular order.
The independent power spectrum components form
the SOAP vector $\mathbf p^{(i)}$.

We use Gaussian type radial basis functions with
$\sigma_{\rm SOAP}=0.25$, $n_{\max}=8$, and
$\ell_{\max}=8$, yielding a 324 component descriptor.
Coslovich et al.~\cite{Coslovich2022JCP} suggest
$\sigma_{\rm SOAP}\approx r_{cut}/n_{\max}$
as a practical guideline for selecting the Gaussian
width, with a width exceeding the typical vibrational
amplitude in thermal systems.

To incorporate information from neighbouring
environments, we average the SOAP vectors in real space:
\begin{equation}
\bar{\mathbf p}^{(i)}
=
\frac{1}{1+N_{\rm nb}(i)}
\displaystyle\sum_{j\in\{i\}\cup N_{\rm nb}(i)}
\mathbf p^{(j)},
\label{eq:soap_local_average}
\end{equation}
where $N_{\rm nb}(i)$ denotes the number of neighbours
within cut off $1.4$ of particle $i$.
Averaging nearby SOAP vectors reduces local
fluctuations~\cite{Donkor2024JPCL,Lionello2025JCTC},
similar to neighbour averaged bond orientational
order parameters, which improve crystal structure
identification~\cite{LechnerDellago2008JCP}.

\subsection{Autoencoder architectures}
\label{sec:ae}

\begin{figure}[t]
    \centering
    \includegraphics[width=\columnwidth]{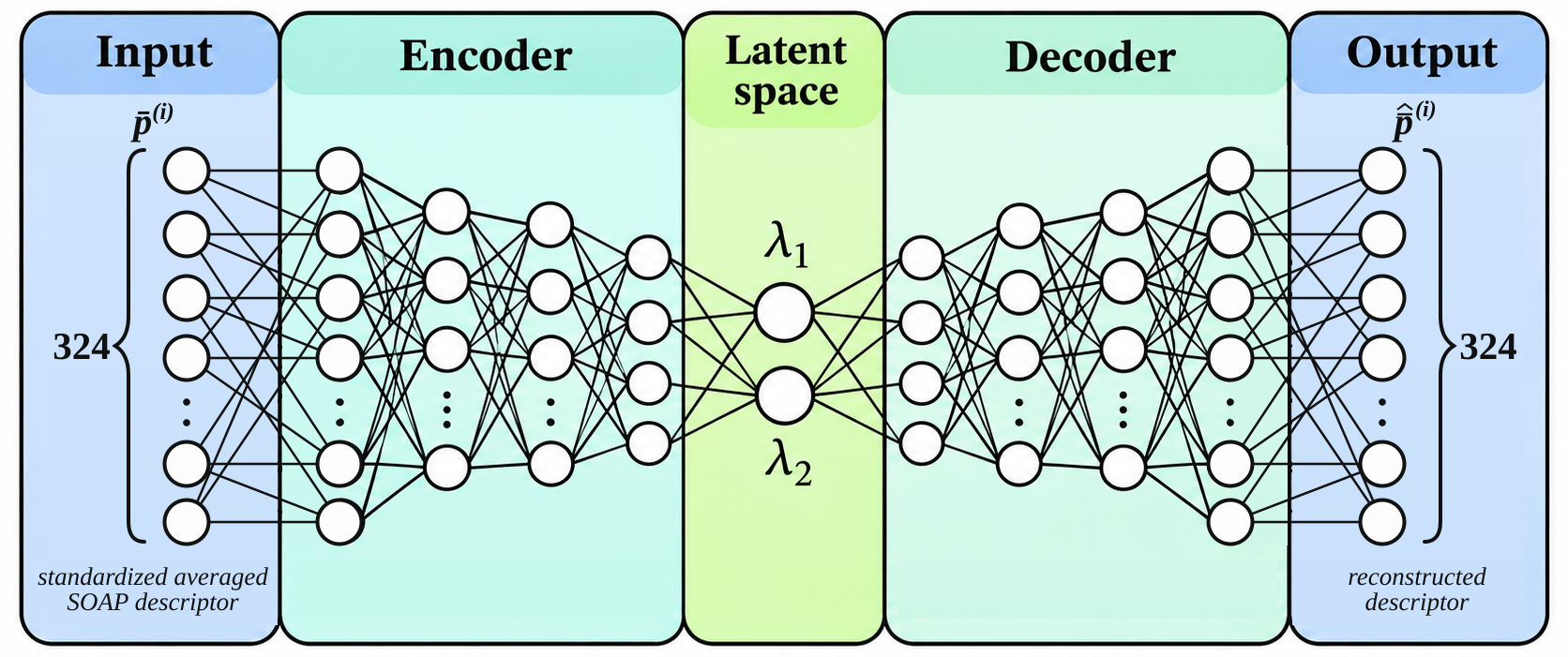}
    \caption{
    Schematic representation of the autoencoder architecture.
    The standardized, locally averaged SOAP descriptor
    $\bar{\mathbf p}^{(i)}$ of dimension $D=324$
    is compressed by the encoder into a two dimensional
    latent representation $(\lambda_1,\lambda_2)$.
    The decoder reconstructs the input descriptor
    as $\hat{\bar{\mathbf p}}^{(i)}$.
    }
    \label{fig:ae_architecture}
\end{figure}

The input to each autoencoder is the locally averaged SOAP descriptor
$\bar{\mathbf p}^{(i)}\in\mathbb{R}^{D}$
of particle $i$, with $D=324$.
The descriptor characterizes the local structural
environment of the particle. Each component is
standardized using its mean and standard deviation,
as described in SI Sec.~S1.

To obtain a low dimensional representation of
the structural descriptors, we use autoencoders,
which learn a compressed representation of the
input data through reconstruction without requiring
predefined structural or dynamical labels.
An autoencoder consists of an encoder
$f_{\theta}:\mathbb{R}^{324}\rightarrow\mathbb{R}^{2}$
with trainable parameters $\theta$, which maps
each descriptor to a two dimensional latent
representation
$\mathbf{z}_i=f_{\theta}(\bar{\mathbf p}^{(i)})
=(\lambda_{1,i},\lambda_{2,i})$,
and a decoder
$g_{\phi}:\mathbb{R}^{2}\rightarrow\mathbb{R}^{324}$
with trainable parameters $\phi$, which reconstructs
the input as
$\hat{\bar{\mathbf p}}^{(i)}=g_{\phi}(\mathbf{z}_i)$.

The two dimensional bottleneck constrains the
network to retain a compressed representation
of the structural information required for
reconstruction. After training, the
encoder is used to obtain the latent space
for subsequent structural analysis.

We compare three autoencoder architectures
that differ in their encoder mappings, while
using the same nonlinear decoder.

\paragraph{Standard autoencoder.}

The Standard AE uses a fully connected
multilayer perceptron (MLP) encoder with
layer widths
$324\rightarrow256\rightarrow256
\rightarrow128\rightarrow2$.
The 324 dimensional input is denoted by
$\mathbf{h}_i^{(0)}=\bar{\mathbf p}^{(i)}$.
The first three transformations
($\ell=0,1,2$) map the input successively
to hidden representations of dimensions
256, 256, and 128. The final
transformation from 128 to 2 dimensions
is linear.

Denoting by $\mathbf{h}_i^{(\ell)}$ the
representation of particle $i$ at layer
$\ell$, the hidden representations are obtained as
\begin{equation}
\mathbf{h}_i^{(\ell+1)}
=
\mathcal{F}^{(\ell)}
\left(\mathbf{h}_i^{(\ell)}\right),
\qquad \ell=0,1,2,
\label{eq:standard_mlp}
\end{equation}
where $\mathcal{F}^{(\ell)}$ denotes the
nonlinear transformation defined as
$\mathcal{F}^{(\ell)}(\mathbf{h})
=
\operatorname{Dropout}_{0.05}
\left[
\operatorname{SiLU}
\left(
\operatorname{LN}
\left(
\mathbf{W}^{(\ell)}\mathbf{h}
+\mathbf{b}^{(\ell)}
\right)
\right)
\right]$.
Here, $\mathbf{W}^{(\ell)}$ and
$\mathbf{b}^{(\ell)}$ are the trainable
weight matrix and bias vector, respectively,
$\operatorname{LN}$ denotes layer normalisation,
and $\operatorname{SiLU}(x)=x/(1+e^{-x})$
is the activation function.
The dropout probability is $0.05$.

The latent coordinates are
obtained from the final hidden representation
through a linear transformation,
$\mathbf{z}_i=
\mathbf{W}_{\rm out}\mathbf{h}_i^{(3)}
+\mathbf{b}_{\rm out}$.

\paragraph{Linear Skip autoencoder.}

The Linear Skip AE uses an affine encoder
mapping from the input SOAP descriptors
to the two dimensional latent space,
\begin{equation}
\mathbf{z}_i
=
\mathbf{W}\bar{\mathbf p}^{(i)}+\mathbf{b},
\label{eq:linear_encoder}
\end{equation}
where
$\mathbf{W}\in\mathbb{R}^{2\times324}$
and $\mathbf{b}\in\mathbb{R}^{2}$ are
trainable parameters. The encoder contains
no nonlinear activation functions, so its
mapping is linear up to the bias term.

\paragraph{Residual autoencoder.}

The Residual AE uses the same encoder widths
as the Standard AE,
$324\rightarrow256\rightarrow256
\rightarrow128\rightarrow2$,
but incorporates a linear skip connection
around each nonlinear hidden block
\cite{He2016ResNet}. The first three
transformations ($\ell=0,1,2$) are residual
blocks, while the final mapping to the
latent space remains linear.

Each residual block is expressed as

\begin{equation}
\mathbf{h}_i^{(\ell+1)}
=
\mathcal{F}^{(\ell)}
\left(\mathbf{h}_i^{(\ell)}\right)
+
\mathbf{S}^{(\ell)}
\mathbf{h}_i^{(\ell)},
\qquad
\mathbf{h}_i^{(0)}=\bar{\mathbf p}^{(i)},
\label{eq:residual_mlp}
\end{equation}

where $\mathcal{F}^{(\ell)}$ consists
of a linear transformation followed
by layer normalisation, SiLU activation,
and dropout with probability $0.05$.
$\mathbf{S}^{(\ell)}$ is  shortcut mapping.

The additive shortcut provides an
additional pathway for information
and gradient propagation through the
network. In contrast to the Standard AE,
the hidden representation is therefore
determined by both the nonlinear
transformation and the linear shortcut.
Setting $\mathbf{S}^{(\ell)}=\mathbf{0}$
reduces Eq.~\eqref{eq:residual_mlp}
to the Standard AE mapping.
The latent coordinates are obtained
through the final linear transformation,
$\mathbf{z}_i=
\mathbf{W}_{\rm out}\mathbf{h}_i^{(3)}
+\mathbf{b}_{\rm out}$.

\paragraph{Decoder.}

The same decoder architecture is used
for the Standard AE, Linear Skip AE,
and Residual AE. It maps the two dimensional
latent representation back to the original
324 dimensional SOAP space through layers
of widths
$2\rightarrow128\rightarrow256
\rightarrow324$.
The intermediate layers use linear
transformations followed by layer
normalisation and SiLU activation,
while the output layer is linear.
No residual shortcuts are included
in the decoder.

\paragraph{Loss function.}

The network parameters are shared across
all particles, and each SOAP descriptor
is processed independently by the
autoencoder. The encoder and decoder
parameters $(\theta,\phi)$ are optimized
by minimizing the mean squared
reconstruction error,
\begin{equation}
\mathcal{L}_{\rm AE}
=
\frac{1}{N_{\rm env}D}
\sum_{i=1}^{N_{\rm env}}
\sum_{k=1}^{D}
\left(
\bar  p_k^{(i)}-\hat{\bar p}_k^{(i)}
\right)^2,
\qquad D=324,
\label{eq:ae_loss}
\end{equation}
where $N_{\rm env}$ is the number of
particle environments included in the
loss evaluation. No dynamical information
is used in the optimization. Dropout is
applied during training and disabled
during evaluation. The training dataset,
preprocessing procedure, optimizer
settings, and model selection criteria
are described in
SI Sec. S1.

\subsection{Mean field caging potential}
\label{sec:caging}

The mean field caging potential quantifies the effective
confinement experienced by a particle due to its surrounding
neighbours. Derived from the Ramakrishnan-Yussouff free energy
functional, it provides a physically motivated measure of the
local structural environment and has been used to investigate
structure dynamics correlations in supercooled
liquids\cite{Sharma2022Softness,mohit_LJ_WCA,Kumawat2025MRO}.
Within this description, the environment surrounding a tagged
particle is treated as frozen, and the dynamics of the
neighbouring particles are neglected.

For a binary mixture, the macroscopic caging potential at zero
displacement is expressed as
\begin{equation}
\beta\Phi =
-\rho \int d^3r
\sum_{u,v} x_u x_v C_{uv}(r)g_{uv}(r),
\label{eq:macro_caging}
\end{equation}

where $\rho$ is the total number density, $x_u$ and $x_v$ are
the mole fractions of species $u$ and $v$, and
$\beta=1/(k_{\mathrm{B}}T)$. Here, $g_{uv}(r)$ and $C_{uv}(r)$
denote the partial radial distribution function and the direct
correlation function, respectively. The partial radial
distribution function is defined as

\begin{equation}
    g_{uv}(r)
    =
    \frac{1}{4\pi r^2\rho_v N_u}
    \left\langle
    \sum_{i\in u}
    \sum_{\substack{j\in v\\j\neq i}}
    \delta_{\mathrm{D}}\!\left(
    r-\left|\mathbf{r}_i-\mathbf{r}_j\right|
    \right)
    \right\rangle,
    \label{eq:partial_rdf_caging}
\end{equation}

where $\rho_v=x_v\rho$, $N_u$ and $N_v$ are the numbers of
particles of the respective species, and
$\delta_{\mathrm{D}}$ denotes the Dirac delta function.
Self pairs are excluded when $u=v$, and the angular brackets
denote an ensemble average. Within the hypernetted chain
approximation, the direct correlation function is given by
\begin{equation}
    C_{uv}(r)
    =
    -\beta U_{uv}(r)+g_{uv}(r)-1-\ln g_{uv}(r),
    \label{eq:hnc_caging}
\end{equation}
where $U_{uv}(r)$ is the pair interaction potential.

To resolve variations in the local environment, the
macroscopic description is extended to individual particles.
The microscopic caging potential for particle $i$ of species
$u$ is written as
\begin{equation}
\beta\phi_u^i =
-\rho\int d^3r
\sum_v x_v C_{uv}^i(r)g_{uv}^i(r),
\label{eq:local_caging}
\end{equation}
where $g_{uv}^i(r)$ and $C_{uv}^i(r)$ are the particle resolved
radial distribution and direct correlation functions. Following
the local normalisation used in
Ref.~\cite{Kumawat2025MRO}, the radial distribution function
around particle $i$ is evaluated by Gaussian smoothing,
\begin{equation}
    g_{uv}^i(r)
    =
    \frac{1}{4\pi\rho r^2}
    \sum_{\substack{j\in v\\j\neq i}}
    \frac{1}{\sqrt{2\pi\delta^2}}
    \exp\!\left[
    -\frac{(r-r_{ij})^2}{2\delta^2}
    \right],
    \label{eq:local_rdf_caging}
\end{equation}
where $r_{ij}$ is the separation between particles $i$ and $j$,
and $\delta=0.09\sigma_{\mathrm{AA}}$ is the Gaussian width.
This smoothing yields a continuous representation of the local
radial structure.

To avoid unphysical short distance contributions
arising from the local application of the
hypernetted chain approximation in
Eq.~\ref{eq:local_caging}, we replace the
direct correlation function $C_{uv}(r)$ with the total
correlation function, $h_{uv}(r)=g_{uv}(r)-1$
\cite{mohit_LJ_WCA,Kumawat2025MRO}. The radial
integration is restricted to the first coordination shell.
For each species pair, the lower integration limit is the
distance at which the ensemble averaged $g_{uv}(r)$ first
becomes nonzero, and the upper limit is the position of its
first minimum.

Although the surrounding particles are treated within a
mean field approximation, the use of particle resolved radial
distribution functions retains the heterogeneity of the local
structure. The resulting microscopic caging potential
therefore provides a particle level structural measure of
confinement by the neighbouring particles.

\section{Results}
\label{sec:results}

\begin{figure*}[t]
    \centering
    \includegraphics[width=\textwidth]
    {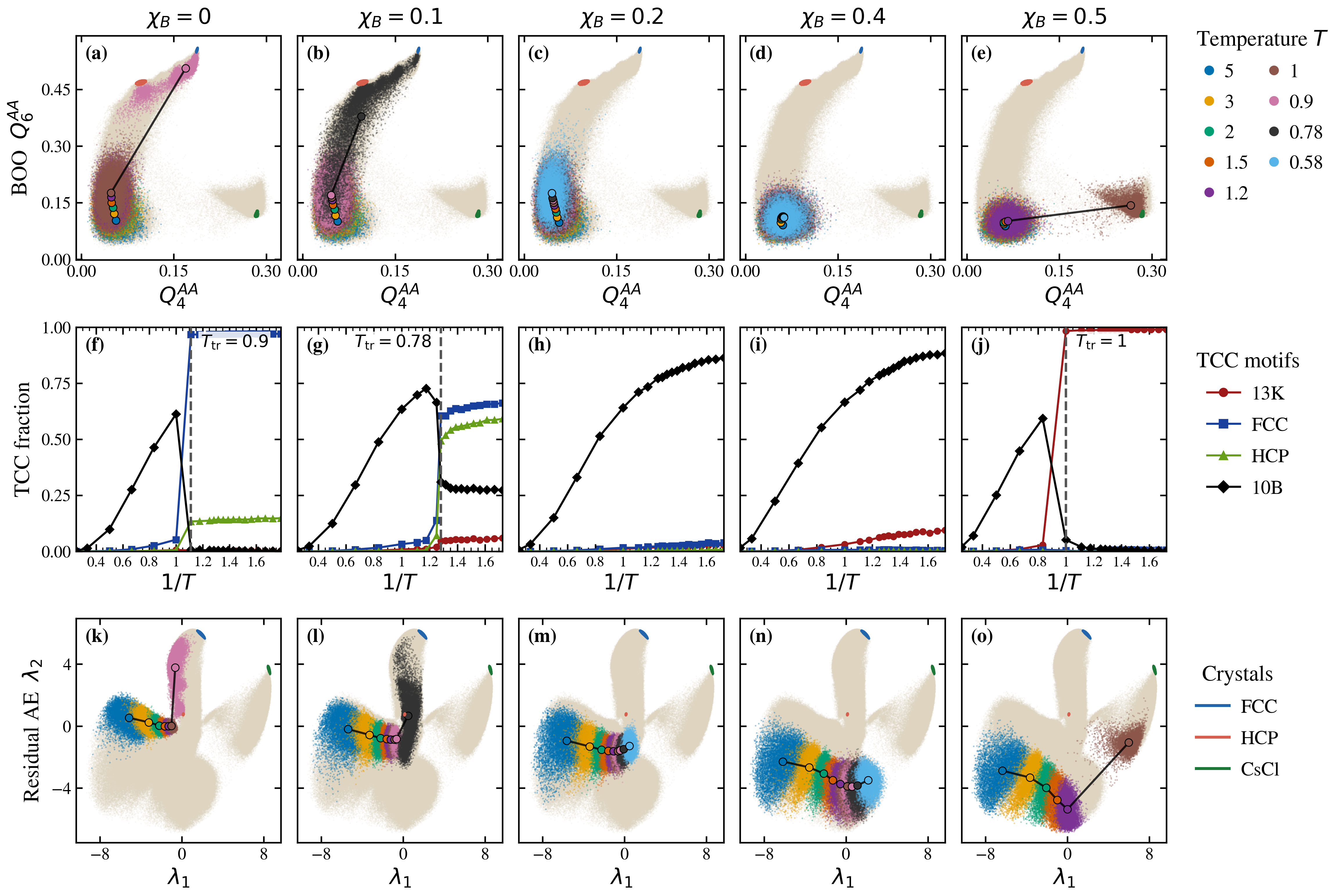}
\caption{Temperature dependent structural evolution for
$\chi_B=0$, $0.10$, $0.20$, $0.40$, and $0.50$ (left to right), from NPT
cooling runs at $P=10.19$. Top row (a--e): distributions of the A-A
bond orientational order parameters in the $(Q_4^{AA},Q_6^{AA})$
plane. Middle row (f--j): temperature dependence of the fraction of
particles belonging to the TCC clusters 13K, FCC, HCP, and 10B  plotted against $1/T$; dashed vertical lines
mark the transition temperature $T_{\rm tr}$ where applicable.
Bottom row (k--o): distributions of A particles in the latent space
$(\lambda_1,\lambda_2)$ of the Residual AE. In the BOO and AE panels, colours denote
temperature, and black circles filled with the respective colours connected by black
lines trace the mean of each distribution. The pale background shows the
pooled reference data from all compositions and temperatures, and the
coloured contours mark the FCC, HCP, and CsCl crystal references.}
    \label{fig:structural_evolution}
\end{figure*}

\begin{figure}[t]
\centering
\includegraphics[
width=\columnwidth,
height=0.72\textheight,
keepaspectratio
]{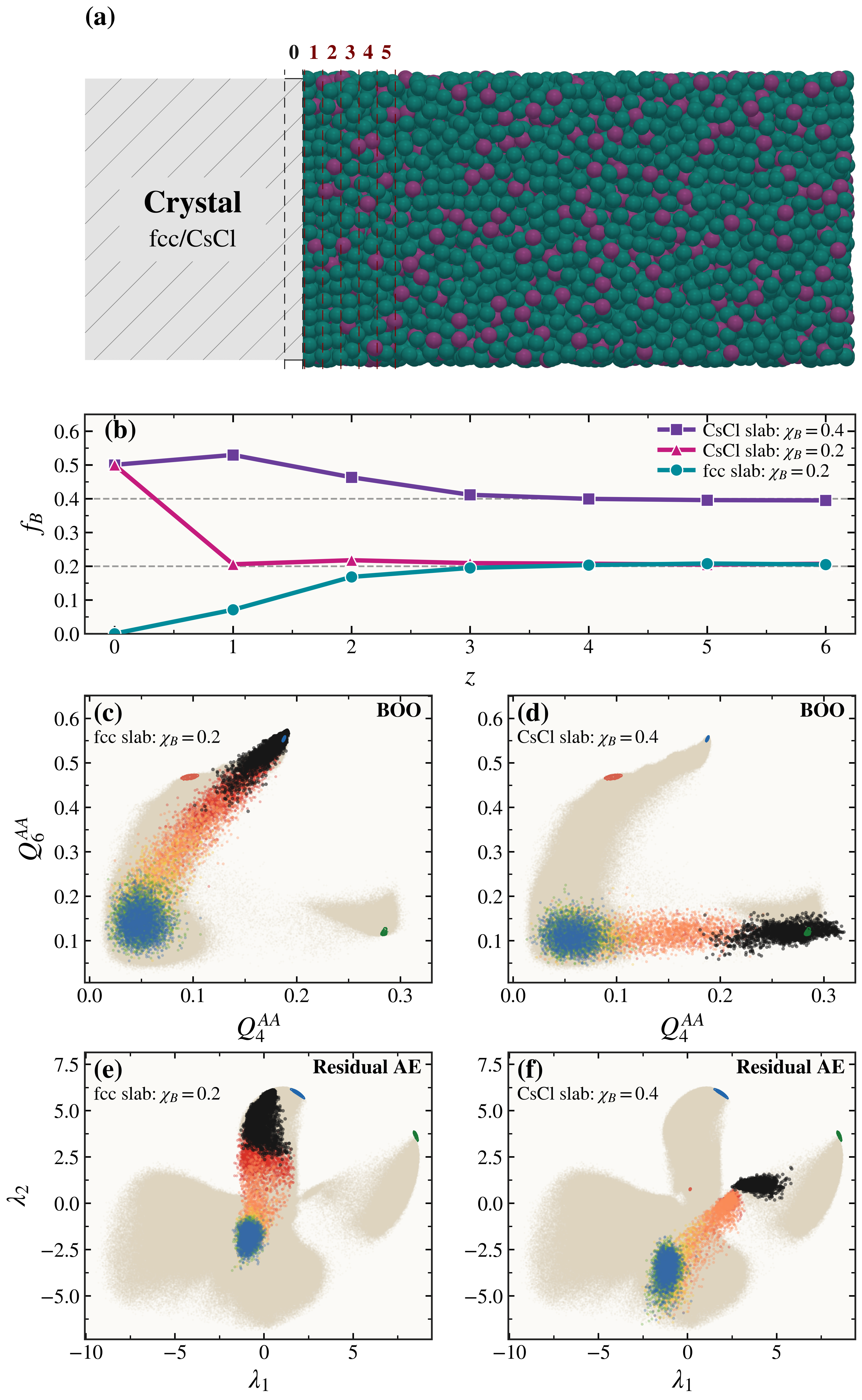}
\caption{Structural variation across crystal liquid interfaces.
(a)~Snapshot of the slab geometry: the pinned crystal (hatched) is in
contact with the liquid, which is divided into slabs numbered $1$--$5$
with increasing distance from the interface; $0$ denotes the crystal slab that faces the liquid.
The liquid shown here has $\chi_B=0.20$. 
(b) B-particle fraction $f_B$ in each slab; dashed lines indicate
the global liquid compositions.
Throughout panels (c--f), black
denotes the crystal, and red, salmon, yellow, green, and blue denote
liquid slabs $1$--$5$, respectively.  Contours mark the FCC (blue), HCP (red), and CsCl (green) crystals.
(c,d)~A--A bond orientational order distributions in the
$(Q_4^{AA},Q_6^{AA})$ plane, shown separately for the particles
in each slab, for the FCC interface at $\chi_B=0.20$ and the CsCl interface
at $\chi_B=0.40$, respectively.
(e,f)~Corresponding slab resolved distributions in the Residual AE latent
space $(\lambda_1,\lambda_2)$.}
\label{fig:slab}
\end{figure}

\begin{figure*}[t]
    \centering
    \includegraphics[width=\textwidth]
    {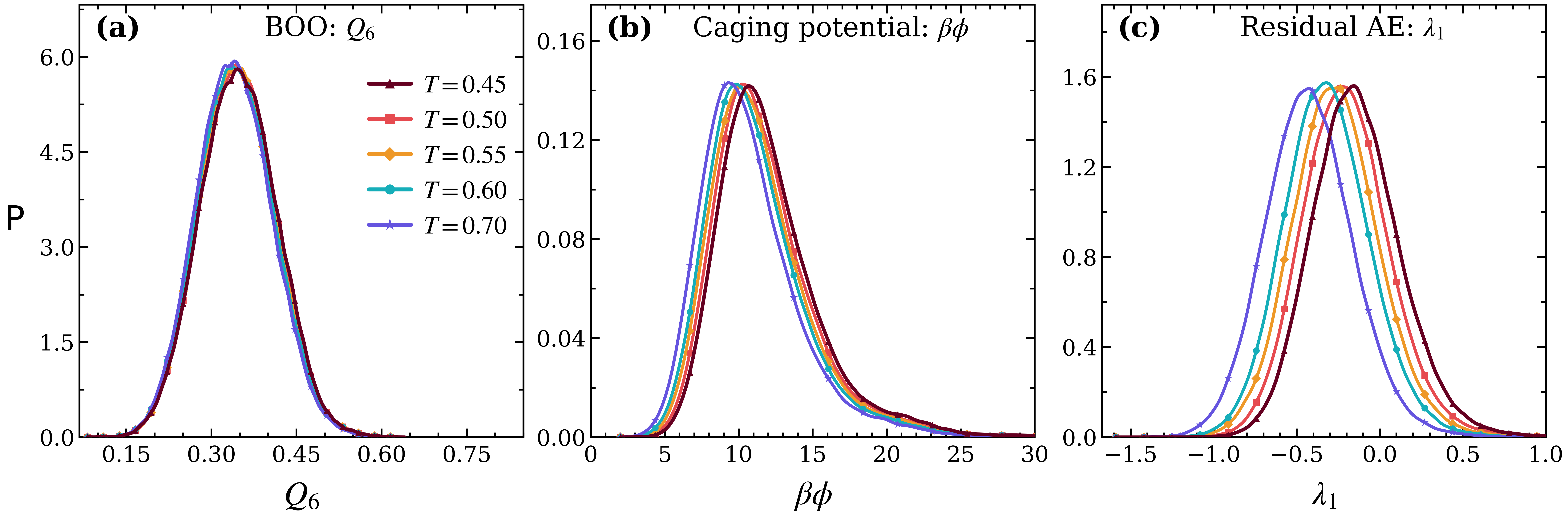}
\caption{Temperature dependent distributions of (a) the bond orientational
order parameter $Q_6$, (b) the caging potential $\beta\phi$, and
(c) the Residual AE coordinate $\lambda_1$ for A particles in the
$\chi_B=0.20$ mixture at $\rho=1.2$ (NVT simulation data). Colours denote temperature, from
$T=0.45$ to $T=0.70$.}

    \label{fig:AE_distribution}
\end{figure*}

\begin{figure*}[t]
    \centering
    \includegraphics[width=\textwidth]  {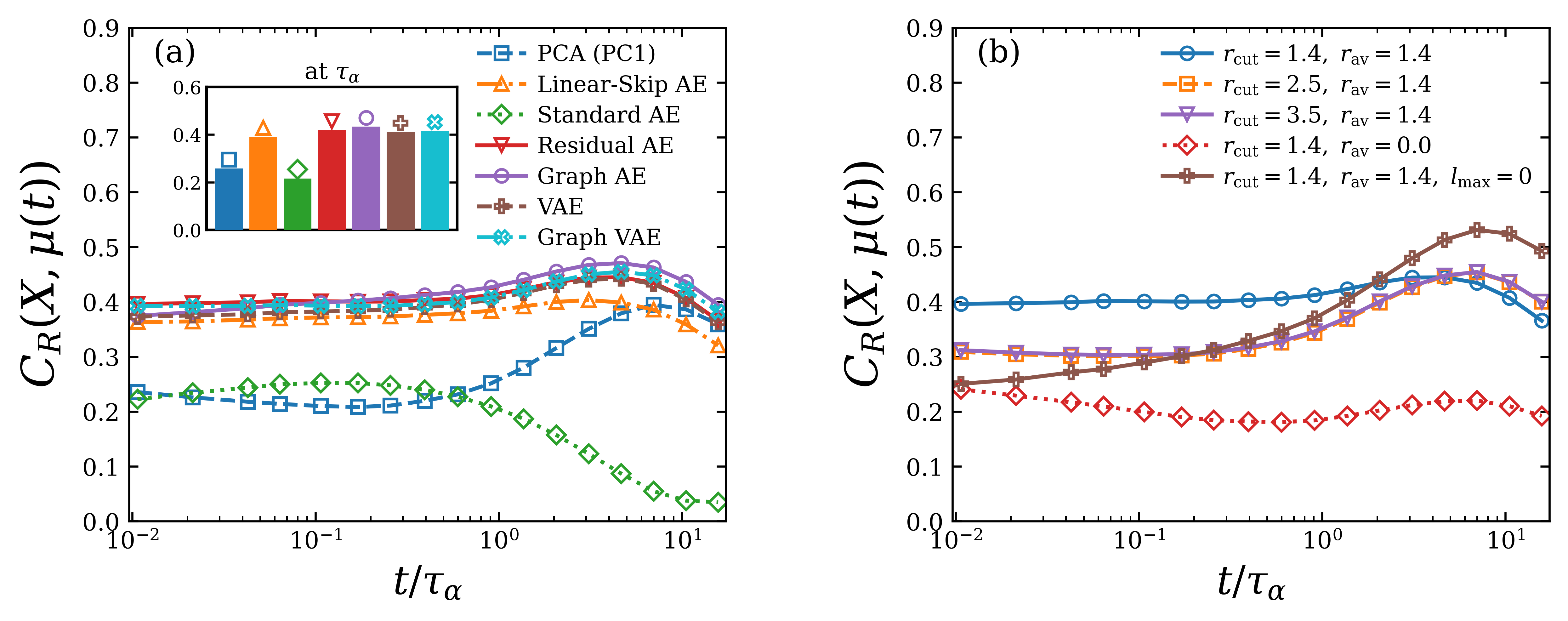}
\caption{Absolute Spearman rank correlation
$C_R(X,\mu(t))$ between structural descriptors
and isoconfigurational mobility $\mu_i(t)$
of A particles in the $\chi_B=0.20$ mixture
at $T=0.45$ and $\rho=1.20$, as a function
of $t/\tau_\alpha$.
(a) Comparison of PCA (PC1), Linear Skip AE,
Standard AE, Residual AE, Graph AE, VAE,
and Graph VAE (both with $\beta_{\max}=10^{-4}$). The inset shows the absolute Spearman rank correlation,
at time $\tau_\alpha$. The colours and symbols
match those of the corresponding models in the main panel.
(b) Comparison of Residual AE models trained
using different SOAP descriptor settings:
$r_{\mathrm{cut}}=1.4$, $2.5$, and $3.5$,
each with neighbour averaging over
$r_{\mathrm{av}}=1.4$;
$r_{\mathrm{cut}}=1.4$ without neighbour
averaging ($r_{\mathrm{av}}=0$);
and radial only SOAP descriptors
($l_{\max}=0$) with $r_{\mathrm{cut}}=1.4$
and $r_{\mathrm{av}}=1.4$.
Correlations are calculated separately
for each configuration, and their absolute
values are averaged over 12 independent
configurations.}
    \label{fig:time_structure_dynamics}
\end{figure*}

\begin{figure}[t]
\centering
\includegraphics[width=\columnwidth]
{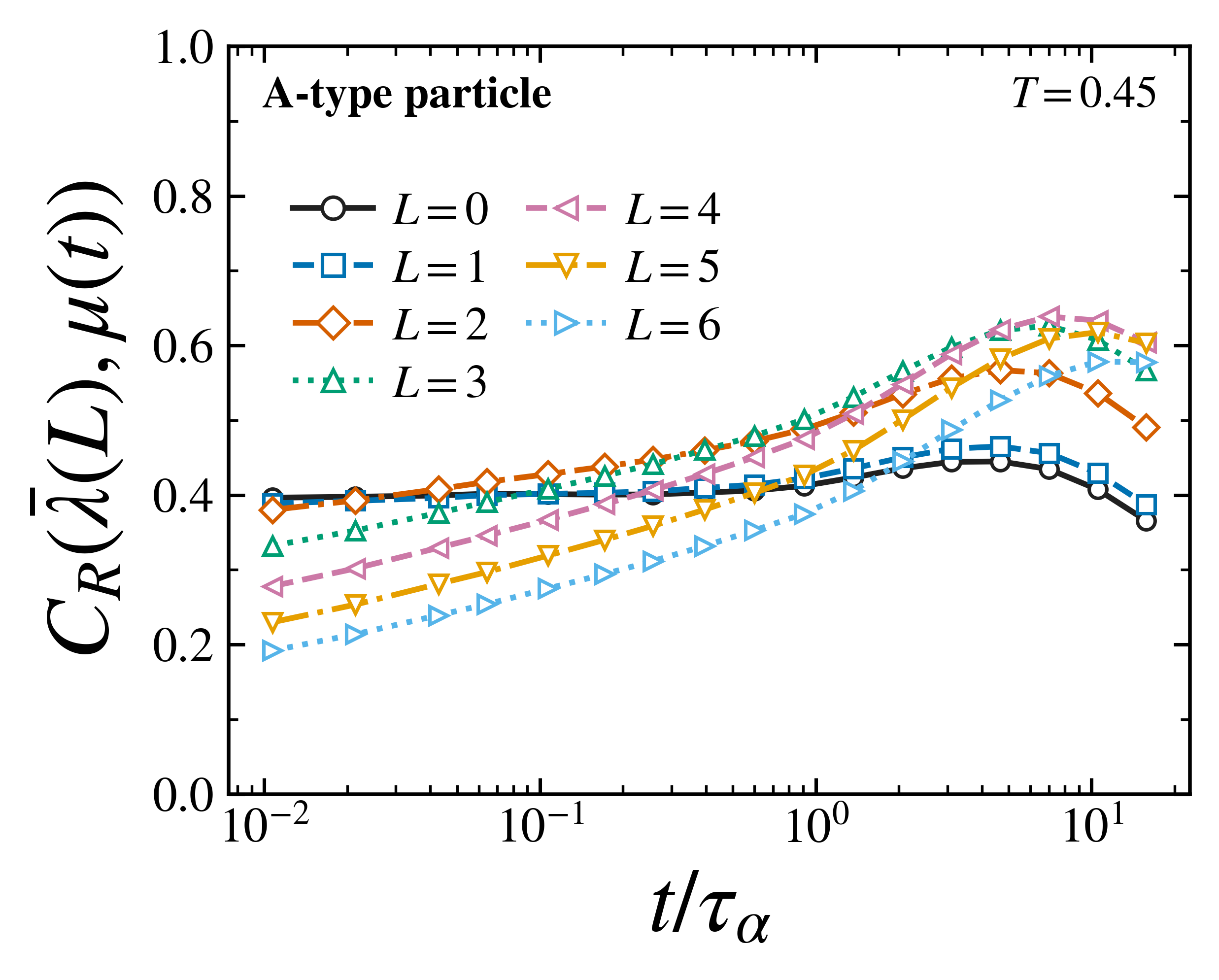}
\caption{Time dependent structure dynamics absolute correlation $C_R(\bar{\lambda}(L),\mu(t))$
as a function of $t/\tau_\alpha$ for the Residual AE coordinate $\lambda_1$ of
A particles in the $\chi_B=0.20$ mixture at $T=0.45$ and $\rho=1.2$.
Different curves correspond to different coarse graining lengths
$L=0$--$6$ (Eq.~\eqref{eq:AE_coarsegrain}). Coarse graining increases the
correlation with long time mobility, and after an optimal length the correlation drops. This optimal length is time dependent.  Correlations are calculated separately for each
configuration, and their absolute values are averaged over 12 independent
configurations.}
\label{fig:AE_time_coarsegrain}
\end{figure}

\begin{figure*}[t]
\centering
\includegraphics[width=\textwidth]
{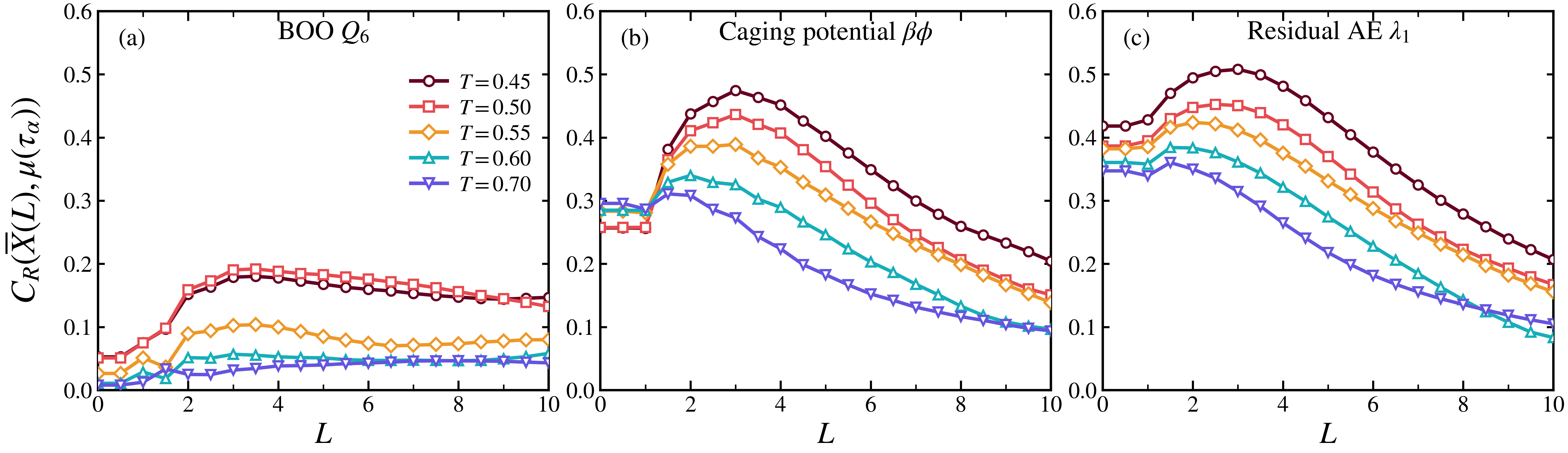}
\caption{Absolute Spearman rank correlation $C_R(\overline{X(L)},\mu(\tau_{\alpha}))$
between the coarse grained structural order parameter and isoconfigurational
mobility $\mu$ at $t=\tau_\alpha$, as a function of the coarse graining length
$L$ (Eq.~\eqref{eq:AE_coarsegrain}), for A particles in the $\chi_B=0.20$
mixture at $\rho=1.2$. The panels show (a)~ $X=Q_6$, the bond orientational order
parameter , (b)~$X=\beta\phi$, the caging potential, and
(c)~$X=\lambda_1$, the Residual AE coordinate. Colours and markers denote
temperatures from $T=0.45$ to $T=0.70$. The caging potential and the
Residual AE coordinate show maxima at a finite coarse graining length
$L_{\max}$ and stronger correlations than $Q_6$.
Correlations are calculated separately for each configuration, and their
absolute mean values are averaged over 12 independent configurations.}
\label{fig:allmodels_coarsegrain}
\end{figure*}

\begin{figure*}[t]
    \centering
    \makebox[\textwidth][c]{%
        \includegraphics[width=1.01\textwidth]
        {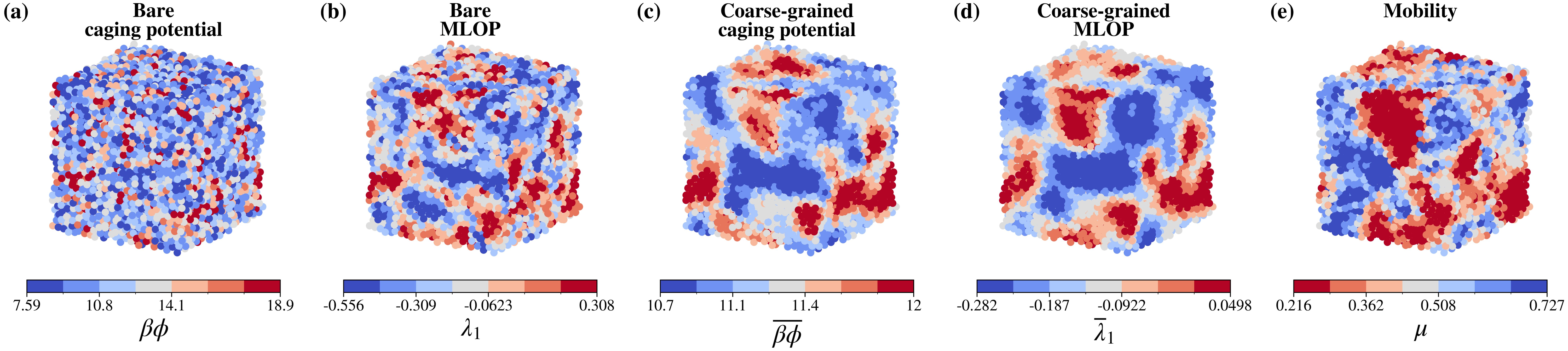}%
    }
\caption{Spatial maps of A particles in the $\chi_B=0.20$ mixture at
$T=0.45$ and $\rho=1.2$, shown at the initial particle positions for the
same configuration: (a)~bare caging potential $\beta\phi$, (b)~bare
Residual AE coordinate $\lambda_1$, (c,d)~their coarse grained values
$\overline{\beta\phi}$ and $\overline{\lambda}_1$ at $L_{max}=3.6$
(Eq.~\eqref{eq:AE_coarsegrain}), and (e)~isoconfigurational mobility $\mu$
at $t=\tau_\alpha$. Each panel uses seven colour bands spanning the 5th to
95th percentiles of its field, with values outside this range assigned
the endpoint colours. The mobility colour map is reversed, so that regions of
low mobility and regions of high $\beta\phi$ or $\lambda_1$ share the same
colour. The bare caging
potential (a) shows no clear patches, whereas the bare Residual AE
coordinate (b) already shows some. After coarse graining (c,d), the
patches of both fields become much clearer and match those of the
mobility in (e). The coarse grained patches of the caging potential and of
the Residual AE coordinate are similar to each other.}
    \label{fig:AE_spatial_maps}
\end{figure*}

\begin{figure}[t]
    \centering
    \includegraphics[width=\columnwidth]
    {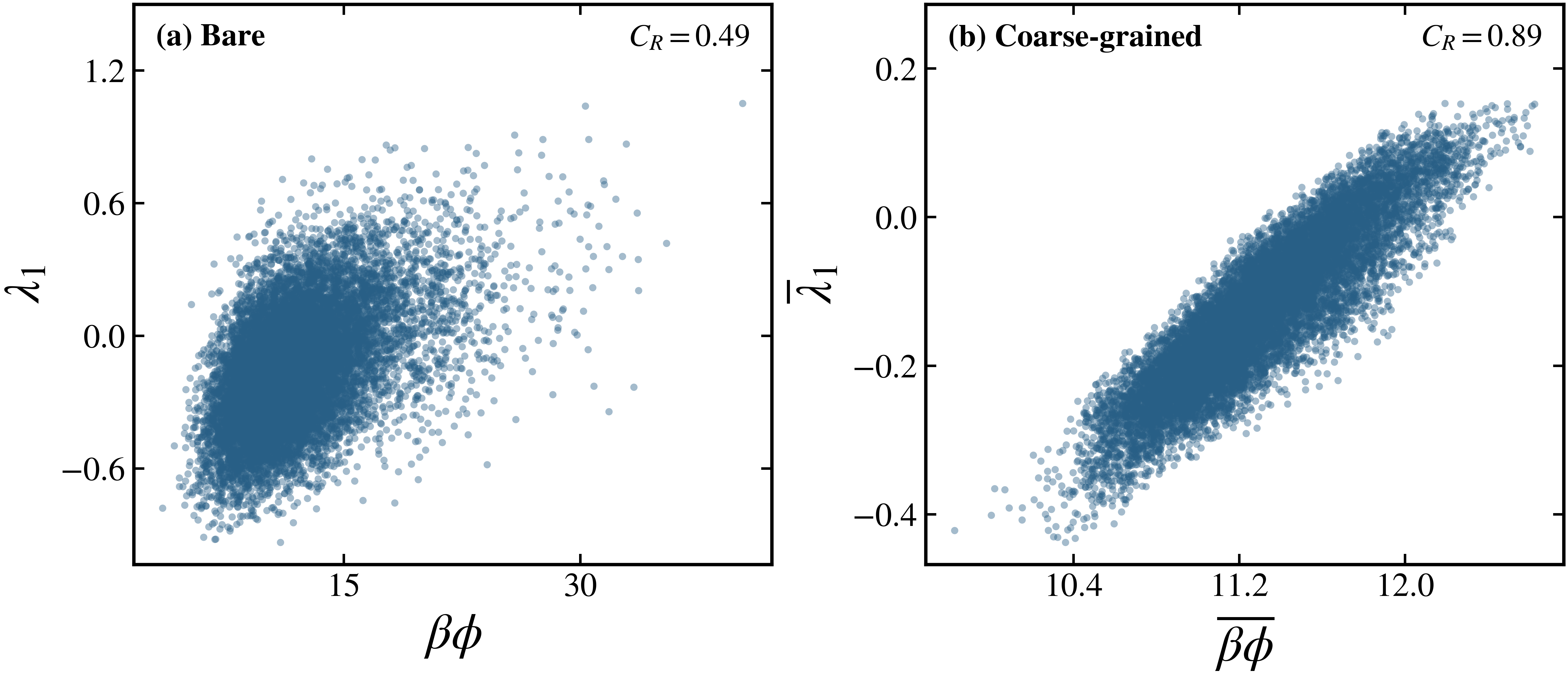}
\caption{
Scatter plots of the Residual AE order parameter
$\lambda_1$ versus the microscopic caging potential
$\beta\phi$ for A particles in a single configuration
of the $\chi_B=0.20$ mixture at $T=0.45$ and
$\rho=1.20$. Each point represents one particle.
(a) Bare quantities $\lambda_1$ and $\beta\phi$.
(b) Their coarse grained counterparts,
$\overline{\lambda}_1$ and $\overline{\beta\phi}$,
evaluated at their respective optimal
coarse graining lengths $L_{\max}=3.6$.
The Spearman rank correlation coefficient
increases from $C_R(\lambda_{1},\beta\phi)=0.49$ for the bare
quantities to $C_R(\overline{\lambda_{1}},\overline{\beta\phi})=0.89$ after coarse graining.
}
 \label{fig:caging_residual_ae_scatter}
\end{figure}

\begin{figure}[t]
\centering
\includegraphics[width=\columnwidth]
{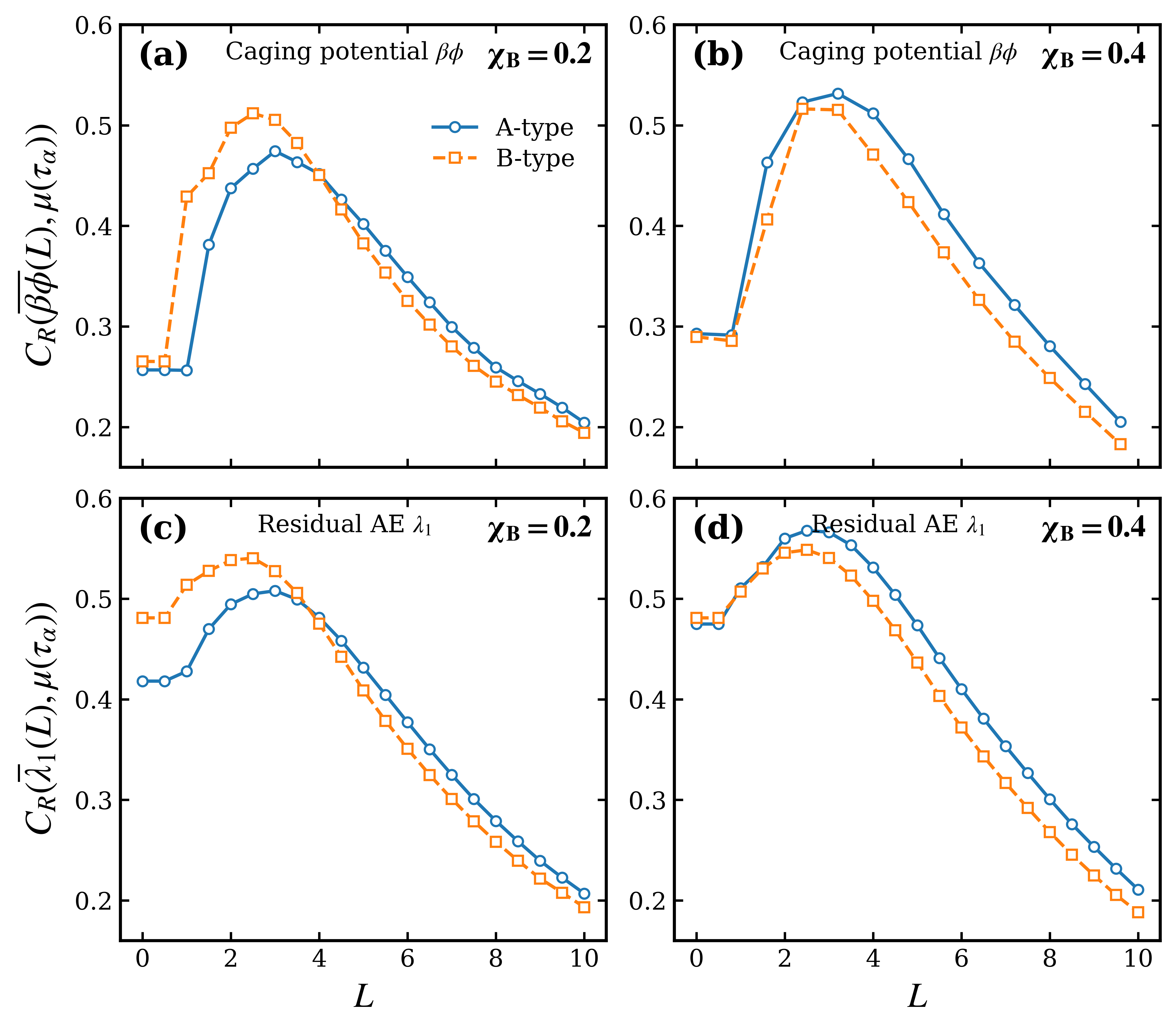}
\caption{Absolute Spearman rank correlation
between the
coarse grained structural descriptors
and isoconfigurational mobility, $\mu$
at $t=\tau_\alpha$, as a function of
the coarse graining length $L$. (Eq.~\eqref{eq:AE_coarsegrain}), at $T=0.45$
for the $\chi_B=0.20$ liquid at $\rho=1.20$ (left column, a,c) and the
$\chi_B=0.40$ liquid at $\rho=1.41$ (right column, b,d). The upper panels
(a,b) show $\beta\phi$, the caging potential, and the lower panels (c,d)
show $\lambda_1$, the Residual AE coordinate. Circles and solid lines denote
A particles; squares and dashed lines denote B particles. For $\chi_B=0.20$ the correlation is higher but peaks at a lower L for B particles than for A particles, whereas for $\chi_B=0.40$ the two species show similar correlations and peak at the same L value. Correlations are
calculated separately for each configuration, and their absolute values
are averaged over 12 independent configurations.}

\label{fig:caging_ae_mkea_composition}
\end{figure}

\subsection{Temperature dependent structural evolution at different compositions}
\label{sec:temperature_evolution}

One of the aims of this paper is to understand the crystalline and precrystalline order in different glass forming systems. For this, we use three different parameter spaces. The first is the standard bond orientational order (BOO) space, traditionally used to identify crystalline structures \cite{Steinhardt1983PRB,LechnerDellago2008JCP}, which is spanned here by ${Q}_{4}$ and ${Q}_{6}$. The second is the Topological Cluster Classification (TCC), which has recently become popular for identifying locally preferred structures in glassy liquids \cite{paddy2013TCC}. The third is a machine learning approach in which an autoencoder (AE) compresses high dimensional SOAP vectors into a two dimensional latent space spanned by $\lambda_{1}$ and $\lambda_{2}$.
The AE models are trained on datasets obtained from molecular dynamics simulations of systems at different compositions $\chi_{B}$ and temperatures. Since some of the systems crystallise at low temperatures, we use NPT simulations at pressure $P=10.19$ \cite{Pedersen2018PRL}. Although we have studied several AE architectures, here we present results for the Residual AE, which gives the best structure dynamics prediction among the models we tested. A comparison of the different AE models is given in Sec.~\ref{subsec:structure_dynamics_overview}.

The results are shown in Fig.~\ref{fig:structural_evolution}. We first discuss the compositions where crystallisation occurs.

For $\chi_{B}=0$, in the BOO space the system undergoes a transition to fcc and hcp like structures, as expected for a monatomic system. In the TCC analysis, the fcc and hcp clusters grow at the temperature at which the transition occurs. The 10B cluster grows on cooling, and its population drops once crystallisation sets in. In the AE space, the system again moves into the hcp and fcc basins at low temperatures.

For $\chi_{B}=0.1$, a similar picture is found in the BOO, TCC and AE spaces. However, since only part of the system transforms into a distorted hcp/fcc structure, the trajectories in the BOO and AE spaces are more spread out. In the TCC analysis, as in $\chi_{B}=0$, the 10B population grows with decreasing temperature until crystallisation sets in, after which it drops while the fcc and hcp populations grow. Because only part of the system crystallises, some 10B clusters survive after the transition, and a smaller fraction of particles show fcc like structure.

For $\chi_{B}=0.5$, the system is known to form the CsCl crystal structure \cite{atreyee_frust,Ujjwal_frust}, which consists of two interpenetrating simple cubic (sc) lattices. In both the BOO and AE spaces, the trajectories clearly evolve into the CsCl basin. In the BOO space, this basin is identified by the formation of an sc lattice among the A particles. In the TCC analysis, the 10B population again grows on cooling until the transition, after which it decreases while the number of particles in the 13K cluster increases. This suggests that the 13K cluster may be associated with CsCl formation. Although the 11A cluster is the locally favoured structure in the $\chi_{B}=0.2$ liquid, we do not trace it as its population is very small, almost negligible\cite{PADDYCrowther2015JCP}. We also do not trace the bcc structure as it is found even in the crystalline state of  $\chi_{B}=0.0$ system where the BOO does not show any presence of bcc.

Crystallisation is known to be frustrated for $\chi_{B}\approx 0.2$--$0.4$ \cite{atreyee_frust,Ujjwal_frust,jeppe_frust}. In agreement with these studies, none of the three methods shows crystal like structures in the $\chi_{B}=0.2$ and $\chi_{B}=0.4$ systems: the trajectories remain in the liquid region of the BOO and AE spaces, and the TCC analysis shows a growth of the 10B population on cooling in both systems.

Although the BOO and Residual AE spaces may appear to give the same picture, they differ in what they resolve. In the BOO space, apart from the systems that clearly transform to a crystal like structure, the trajectories for all other compositions and temperatures nearly overlap. In the Residual AE space, by contrast, the trajectories at each temperature and composition are well separated. SOAP contains not only the orientational information used in the BOO parameters but also radial information, and we attribute the separation of the different temperatures and compositions primarily to this radial information. Also note that in the SOAP input, we do not differentiate between the two species (See SI Sec. S1). Thus, the composition information is not directly fed in the input. However, the model can separate the compositions.
The Residual AE space therefore retains more detailed information about the local structure of the liquid than the BOO space.

We do not include any reference crystal structures in the training set, and no crystal labels are used. The training data do contain trajectories in which a transition to a crystalline structure occurs, wherever applicable, and this is sufficient for the model to identify the crystal basins.

\subsection{Crystallization tendencies at $\chi_B=0.20$ and $0.40$}
\label{sec:slab}

In an earlier study involving some of us, biased Monte Carlo (MC) simulations showed that, although no crystal like structures are observed in the $\chi_{B}=0.2$--$0.4$ liquids (as also seen in the previous section), the $\chi_{B}=0.2$ system has a tendency to form fcc, while the $\chi_{B}=0.4$ system has a tendency to form CsCl \cite{Ujjwal_frust}. Here we study these crystallisation tendencies, as seen in the BOO and Residual AE spaces, at the crystal liquid interface for the $\chi_{B}=0.2$ and $\chi_{B}=0.4$ systems.

To this end, we create crystal liquid interface systems by placing a crystalline slab in contact with the corresponding liquid, following Ref.~\cite{Ujjwal_frust}. The most stable crystalline state for $\chi_{B}=0.2$ is a coexistence of a CsCl crystal, formed by A and B particles, and an fcc crystal formed by the rest of the A particles, so this system may have a tendency to form both crystal structures \cite{FernandezHarrowell2003PRE}. For $\chi_{B}=0.2$ we therefore study the liquid at the interface with a CsCl crystal and, separately, with an fcc crystal. For $\chi_{B}=0.4$, a CsCl crystal slab is placed in contact with the liquid. The crystalline particles are kept fixed by pinning, while the liquid particles evolve freely. All the crystal liquid interface systems contain $24{,}000$ particles, of which $4{,}000$ are in the crystalline region and $20{,}000$ in the liquid region.  The simulation box is elongated along $z$, with initial dimensions $L_x=L_y=15.496\,\sigma_{AA}$ and $L_z=90.393\,\sigma_{AA}$, and periodic boundary conditions in all three directions. Simulations are performed in the $NP_zT$ ensemble at $P_z=10.19$ and $T=1.1\,T_m$, where $T_m$ is the composition dependent melting temperature reported in Ref.~\cite{Pedersen2018PRL}. This choice of temperature is to make sure that no liquid particles crystallise on the crystal surface.

Crystallisation requires the local composition to be close to that of the crystal, so we first track the fraction of B particles, $f_{B}$, in slabs parallel to the interface. Each slab is of $1 \sigma_{AA}$ width. For $\chi_{B}=0.2$ at the interface with the CsCl crystal, $f_{B}$ near the interface is the same as in the bulk liquid. Thus, even in the presence of the crystal slab, there is no composition change that would eventually favour CsCl formation. This is consistent with the earlier biased MC study, in which even after bias, a CsCl crystal did not form in the $\chi_{B}=0.2$ liquid \cite{Ujjwal_frust}. When the $\chi_{B}=0.2$ liquid is placed next to the fcc crystal, $f_{B}$ is depleted near the interface and reaches the bulk value only at the fourth slab. Similarly, when the $\chi_{B}=0.4$ liquid is placed next to the CsCl crystal, $f_{B}$ is enhanced near the crystal and reaches the bulk value only around the fourth slab. This shows that the liquid near the fcc crystal at $\chi_{B}=0.2$, and near the CsCl crystal at $\chi_{B}=0.4$, has a local composition similar to that of the corresponding crystalline phase.

We next study these two systems in the BOO and Residual AE spaces. We project the particles of the zeroth (crystal slab in black) and the first 5 slabs of the liquid onto these spaces, using a different colour for each slab. In both spaces, the first slab lies closest to the corresponding crystal structure, and as we move away from the crystal, the particles move away from the crystal basins. In the BOO space, the particles of the zeroth slab appear to have a perfect crystal structure, whereas in the Residual AE space they lie near the crystal basin but remain displaced from the perfect crystal. 
This suggests that the orientational order is crystal like, while the radial distances may not match those of the crystals. 
Note that this crystal slab faces the liquid, and in the feature vector space, where information from the neighbours is needed, the liquid particles are on one side. This might distort the description of the perfect crystal structure, which is shown in the residual AE space, suggesting that the residual AE space can differentiate smaller changes in the structure.

\subsection{Structure dynamics correlation}
\label{subsec:structure_dynamics_overview}

Having established that our machine learned order parameter (MLOP) is a good descriptor of the crystallisation tendencies of these systems, we now use it to study the correlation between structure and dynamics. For this part of the work, we project configurations of NVT simulations of the $\chi_{B}=0.2$ and $\chi_{B}=0.4$ systems onto the latent space learned from the NPT simulations.
The dynamics of supercooled liquids is heterogeneous, and this heterogeneity grows as the temperature decreases. The question we address is how much of it is attributable to structural differences.

One important point is that in the latent space $\lambda_{1}$ is the direction which primarily shows the temperature dependence of a system, whereas crystallisation in the $\chi=0.2$ is shown by movement in the $ \lambda_{2} $ direction (Fig.\ref{fig:structural_evolution}). We will show that in the liquid state $\lambda_{1}$ is a better predictor of the dynamics. Thus, our study shows that in this system it is the finer structural difference in the liquid and not the crystal like structure that predicts the dynamics. This confirms earlier studies where crystal like structure and its growth were found to be absent in this system and uncorrelated with slow dynamics \cite{Ujjwal_frust,atreyee_frust,
PADDYCrowther2015JCP,FernandezHarrowell2003PRE}.

In Fig.~\ref{fig:AE_distribution} we plot the temperature dependence of the distributions of three structural order parameters, averaged over 12 initial configurations: the bond orientational parameter $Q_{6}$, the caging potential $\beta\phi$ (Sec.~\ref{sec:caging}), obtained from the microscopic theory, and the MLOP $\lambda_{1}$. All three show broad distributions. However, the temperature dependence of $Q_{6}$ is negligible, whereas $\lambda_{1}$ shows a significant temperature dependence, as discussed earlier.

The structure dynamics correlation is studied using the isoconfigurational ensemble (Sec.~\ref{sec:mobility}) introduced by Harrowell and co-workers \cite{WidmerCooper2004PRL} to separate the influence of the initial structure from that of thermal noise in the dynamics. In a single trajectory, the displacement of a particle depends on both its local structure and its randomly assigned Maxwell Boltzmann velocity, which can mask the structural effect. We therefore generate many trajectories from the same initial configuration, each with independently sampled velocities. Averaging over this ensemble suppresses the thermal noise, so that the isoconfigurational average mobility $\mu_i(t)$ (Eq.~\eqref{eq:mobility}) reflects the influence of the initial structure. The A and B particles are analysed separately to obtain cleaner statistics.

To quantify the relationship between local structure and the particle mobility, we use the Spearman rank correlation coefficient. In the absence of tied ranks, it is given by
\begin{equation}
C_R(X,Y)
=
1-\frac{6\sum_{i=1}^{n}d_i^2}{n(n^2-1)},
\label{eq:bare_time_corr}
\end{equation}
where $d_i$ is the difference between the ranks of $X$ and $Y$ for particle $i$, and $n$ is the number of particles considered. Here $X$ is the structural order parameter, for example $\lambda_1$, $\beta\phi$ or $Q_6$ and $Y$ is the mobility $\mu_i(t)$. Correlations are computed separately for A and B particles in each configuration, and their absolute values are then averaged over $12$ independent configurations.

Here we first present the comparative predictive power of the different AE models. Figure~\ref{fig:time_structure_dynamics}(a) shows the results obtained from PCA (PC1) and the different AE architectures: the Linear Skip AE, the Standard AE (a plain MLP autoencoder), the Residual AE, the Graph AE, the VAE and the Graph VAE, where the same SOAP parameters across compositions and temperatures are provided during training. The Linear Skip AE uses a linear encoder, whereas the Standard AE uses a nonlinear encoder without a linear skip connection. The Residual AE combines a linear skip connection with nonlinear transformations. 
The Graph AE, implemented using a Graph Attention Network (GAT), incorporates information from neighbouring particle environments through graph message passing, with learned attention weights
determining the relative contributions of
neighbouring particles~\cite{Velickovic2018GAT}.
 The VAE learns a probabilistic latent representation by minimising a loss that combines the reconstruction error with a regularisation of the latent distribution, and the Graph VAE extends the graph based encoder to this variational framework.

Among the models shown, the Residual AE gives stronger mobility correlations than the Standard AE. The Linear Skip AE yields appreciable mobility correlations. 
The Residual AE and Graph AE give
correlations of comparable magnitude over part of
the investigated time range.
 Therefore, in the rest of the study we use the MLOP obtained from the Residual AE architecture, which gives the strongest correlations among the models tested without requiring a graph architecture. The VAE and Graph VAE show good structure dynamics correlations, generally lower than those of the Residual AE.

In an earlier study using a time lagged autoencoder (TAE), it was shown that the structure dynamics correlation has contributions from particles beyond the first neighbours \cite{QiuYethiraj2025PNAS}. In Fig.~\ref{fig:time_structure_dynamics}(b) we examine how the descriptor affects the correlation obtained with the Residual AE. We vary the SOAP cutoff, test the effect of locally averaging the SOAP descriptors (Sec.~\ref{sec:soap}), and test the effect of discarding the orientational information by setting $l_{\max}=0$. We find that at the $\alpha$ relaxation time the correlation is highest when the SOAP vectors are locally averaged (Eq.~\eqref{eq:soap_local_average}), when both radial and orientational information are included, and when $r_{\mathrm{cut}}=1.4$. Increasing $r_{\mathrm{cut}}$ beyond this value does not improve the correlation in the range of the $\alpha$ relaxation time, $\tau_{\alpha}$. Thus, in the residual AE, unlike in the TAE study, a longer SOAP cutoff does not improve the correlation in our model.


As mentioned before, all the models are trained on structural data from NPT simulations at $P=10.19$ and evaluated on independent NVT configurations. This tests the transfer of the learned representation between simulation conditions.

We find that the structure dynamics correlation decreases over long times. As discussed in earlier studies, the structural environment of a single particle describes its dynamics only up to a certain time, whereas the dynamics over longer timescales also depend on the environments of its neighbours. As proposed by Tanaka and coworkers \cite{TongTanaka2018PRX,TongTanaka2019NatCommun}, and used in other studies \cite{SharmaOzawa2024JCP,Kumawat2025MRO}, the effect of the neighbours can be included by coarse graining the order parameter. The coarse graining of an order parameter $X$ over a length $L$ is performed as
\begin{equation}
\overline X_i(L)
=
\frac{
X_i+
\displaystyle\sum_{\substack{j\ne i\\r_{ij}\le L}}
X_j\exp(-r_{ij}/L)
}{
1+
\displaystyle\sum_{\substack{j\ne i\\r_{ij}\le L}}
\exp(-r_{ij}/L)
}.
\label{eq:AE_coarsegrain}
\end{equation}
Here $r_{ij}$ is the minimum image separation in the initial configuration, and the sum includes both species. The central particle contributes with unit weight, and the case $L=0$ is defined separately as $\overline X_i(0)=X_i$. Note that the importance of coarse grained order parameters is also predicted by ML studies \cite{JungBiroliBerthier2024PRB,
SharmaOzawa2024JCP}.

In Fig.~\ref{fig:AE_time_coarsegrain} we plot the Spearman rank correlation between the coarse  grained latent parameter $\lambda_{1}(L)$ in the Residual AE space and the mobility, $\mu(t)$ at different times for different coarse graining lengths, $L$ at a temperature $T=0.45$. We find that at short times, the correlation is high for smaller coarse graining lengths, whereas at long times the correlation initially grows with coarse graining length and beyond a certain optimal length it drops. Note that the optimal length increases with time. This supports the earlier argument that in predicting the long time dynamics it is not only the structural information of the particle but its neighbours' structural information is also needed. 

Next, we study the temperature variation of the Spearman rank correlation and the optimal length that is obtained at the fixed value of time, the $\alpha$ relaxation time, $\tau_{\alpha}$. For that, in Fig.~\ref{fig:allmodels_coarsegrain} we plot the Spearman rank correlation between the coarse grained order parameters and the mobility at $\tau_{\alpha}$, as a function of the coarse graining length $L$, at different temperatures. The non monotonic $L$ dependence of the correlation, discussed above, is clearly observed in this plot, with the structure dynamics correlation showing a maximum at an optimal length $L=L_{\mathrm{max}}$. 
For comparison, we also plot the correlations between the BOO($Q_{6}$) and the mobility and also the $\beta \phi$ and the mobility. The correlation between ${Q}_6$ and the dynamics is very poor. The correlations of the caging potential, $\beta \phi$ and of the MLOP, $\lambda_{1}$, with the dynamics are very similar, with the latter slightly higher. 

The improvement of the correlation with coarse graining can be visualised in Fig.~\ref{fig:AE_spatial_maps}, where we show heat maps of the bare, $L=0$  and the coarse grained, $L=L_{\mathrm{max}}$, caging potential and the latent parameter in the residual AE space, together with the mobility map at $\tau_{\alpha}$ for $T=0.45$. The patches of both order parameters match those of the mobility only after coarse graining, which accounts for the higher Spearman rank correlation. 

Both the $L_{\mathrm{max}}$ and the correlation at $L_{max}$ increase as the temperature decreases, so the effect of coarse graining depends on temperature and time. As the temperature decreases, structure contributes more to the dynamics \cite{andrea_NatPhys2016,Sharma2022Softness}, which increases the correlation. Also, at lower temperatures, the size of the cooperatively rearranging regions grows \cite{CRR_AdamGibbs1965}, and the dynamics of a particle becomes coupled to that of more distant particles. This is consistent with the increase of $L_{\mathrm{max}}$ on cooling. 

\begin{figure*}[t]
    \centering
    \includegraphics[
        width=\textwidth,
        trim={0 0.25cm 0 0},
        clip
    ]{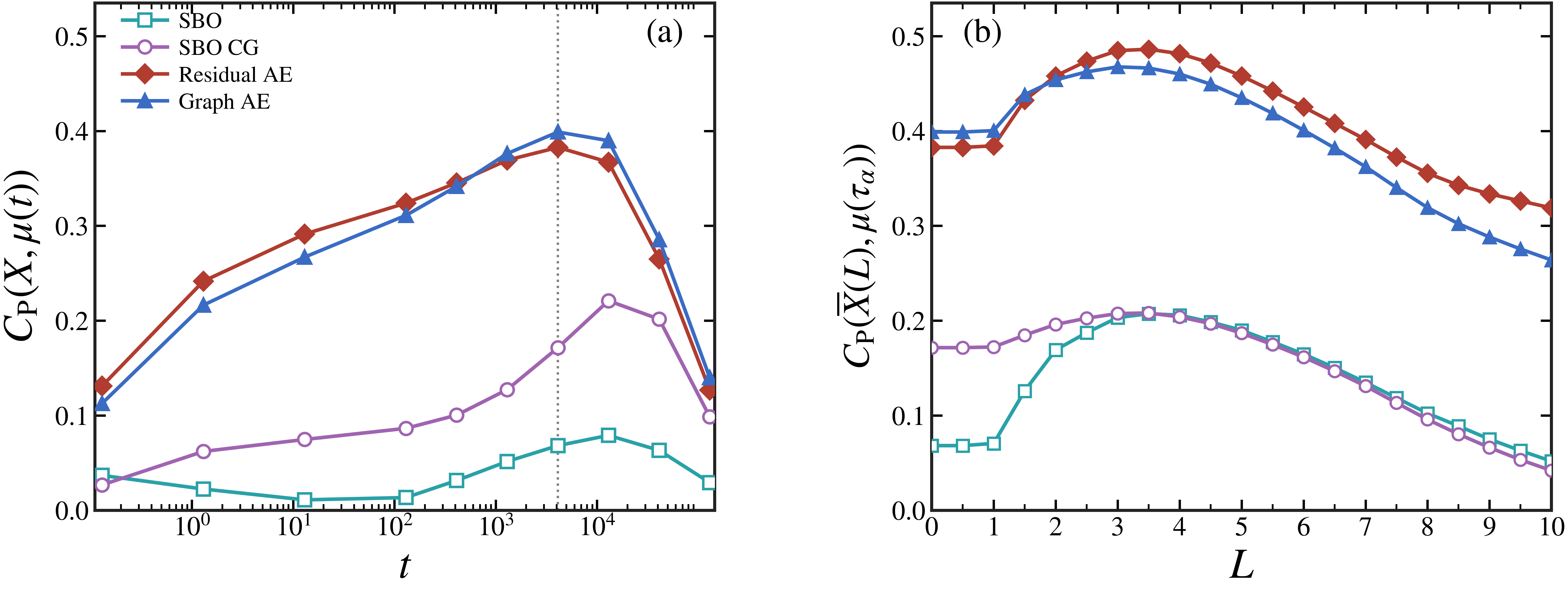}
    \vspace{-0.25cm}
    \caption{Comparison of structure dynamics correlations for the
    three dimensional Kob Andersen binary mixture at
    $T=0.44$, using A particles from 100 configurations
    of the GlassBench dataset \cite{Jung2025Roadmap}.
    (a) Absolute Pearson correlation between particle level
    structural descriptors $X$ and isoconfigurational
    dynamic propensity $\mu(t)$ as a function of time $t$.
    The vertical dotted line indicates the structural
    relaxation time $\tau_\alpha$.
    (b) Absolute Pearson correlation between the
    coarse grained structural descriptors $\overline{X}(L)$
    and dynamic propensity, $\mu$ at $t\simeq\tau_\alpha$,
    as a function of coarse graining length $L$.
    We have followed the same coarse graning protocal as mentioned in Sec.~\ref{subsec:structure_dynamics_overview},
     SBO and SBO CG are taken from the GlassBench dataset,
    whereas Residual AE and Graph AE represent the first
    latent coordinates obtained from the configuration coordinates available from the GlassBench data converted into SOAP vectors and projected on our pretrained models.
    Both learned descriptors exhibit stronger correlations
    with particle dynamics than the SBO based descriptors,
    with Graph AE showing the highest correlation at $\tau_{\alpha}$. Since in the SBO and SBO CG data only A particle information is provided, for this figure in all the coarse graining process we have neglected the B particles.
    }
    \label{fig:glassbench_pearson}
\end{figure*}

At all temperatures at $L=0$ (no coarse graining), the correlation between $\lambda_{1}$ and mobility is higher than that between $\beta\phi$ and mobility as shown in the Fig. \ref{fig:allmodels_coarsegrain} and SI Fig. S4 and S5.  This can also be visualised in Fig.~\ref{fig:AE_spatial_maps}, where at $L=0$ the patches of $\lambda_{1}$ are larger than those of the $\beta \phi$. Note that the SOAP vectors here are locally averaged, whereas such a kind of local averaging is not present in the calculation of bare caging potential. Earlier studies have shown that coarse graining of a structural order parameter improves its correlation with long time dynamics\cite{SharmaOzawa2024JCP, Jung2023PRL,TongTanaka2018PRX,Kumawat2025MRO, Boattini2021PRL}. Interestingly, as shown in Fig \ref{fig:time_structure_dynamics}, for the Graph AE the correlation at L=0 is marginally higher than that predicted by residual AE. However, we find that the coarse graining of the Graph AE does not improve the correlation as much as it does for the residual AE (See Fig: \ref{fig:glassbench_pearson} and Fig. S6 in SI). This is because, as discussed earlier, the Graph architecture, through its connectivity, inherently incorporates some coarse graining, and the improvement from further coarse graining is less effective \cite{Boattini2021PRL}.

In the SI we present the structure dynamics correlation for the B particles in the $\chi_{B}=0.2$ system and for the A and B particles in the $\chi_{B}=0.4$ system. The observations are similar to those for the A particles in the $\chi_{B}=0.2$ system: $\bar{q}_6$ is a poor predictor of the dynamics, the caging potential and the MLOP perform similarly, and at $L=0$ the MLOP performs better. Our generalised ML model therefore predicts the structure dynamics correlation equally well for different compositions.

These results show that, although the MLOP performs better, its predictions are very similar to those of the microscopic theory. We also find a good Spearman rank correlation between the MLOP and the microscopic caging potential, which is about $0.49$ for the bare values and increases to $0.89$ when both order parameters are coarse grained to $L_{\mathrm{max}}=3.6$  (Fig.~\ref{fig:caging_residual_ae_scatter}). Together, these results suggest that the information captured by the ML model is closely related to that of the microscopic caging potential. The MLOP shows a slightly higher correlation with the dynamics, which we attribute to the additional orientational information contained in the SOAP input.

We next discuss further similarities between the ML order parameter and the microscopic caging potential in terms of their predicting the dynamics. In Fig.~\ref{fig:caging_ae_mkea_composition} we compare the structure dynamics correlation of the A and B particles, as predicted by the MLOP and by the caging potential. This comparison is performed for both the $\chi_{B}=0.2$ and $\chi_{B}=0.4$ systems at the lowest temperatures studied. We find that both order parameters make similar predictions. In the $\chi_{B}=0.2$ system, the structure dynamics correlation is higher and peaks at a lower L for B particles than for  A particles, whereas in the $\chi_{B}=0.4$ system it is similar for both species.

A previous study using the caging potential showed that the structure dynamics correlation is higher in an attractive system than in a repulsive one \cite{mohit_LJ_WCA}, which suggests that the correlation increases with the strength of the attraction. In the binary systems that we study here, the attraction between an A and a B particle is higher than that between two A particles. In the $\chi_{B}=0.2$ system, the B particles are primarily surrounded by A particles, with which they interact more strongly, whereas the A particles are primarily surrounded by other A particles, where the interaction is weaker. The local cage, due to its larger effective interaction with the neighbours, is therefore expected to contribute more to the dynamics of a B particle. The scenario is different in the $\chi_{B}=0.4$ system: owing to the larger number of B particles, both species have similar neighbourhoods, and the local caging potential contributes similarly to the dynamics of both.  
Interestingly, although no information about the interactions or the dynamics is fed into the ML model, the MLOP shows the same trend as the caging potential. As discussed above, since the MLOP includes both radial and orientational information, its correlation at $L=0$ is higher than that of the microscopic theory.


\section{Comparison with existing models} 
\label{sec:comparison}

We next discuss the similarities and differences of our study with the existing ML studies for the $\chi_{B}=0.2$ Kob Andersen Model. 
Note that the ML studies include both supervised \cite{andrea_NatPhys2016,Bapst2020NatPhys,
Alkemade2022JCP,Shiba2023JCP,Jung2023PRL} and unsupervised \cite{Boattini2020NatCommun,Coslovich2022JCP,
QiuYethiraj2025PNAS} learning. Although supervised models generally achieve higher predictive accuracy, unsupervised approaches have complementary advantages: they require no dynamical data, they can in principle be trained once and applied to other conditions as we show here, and the learned coordinates can be examined for physical content, as we do here by comparing the MLOP with the microscopic caging potential. Thus, here we compare our results with studies that use unsupervised learning methods \cite{QiuYethiraj2025PNAS,Boattini2020NatCommun}.

In ML studies of the KA model, there are many variations in how input data are provided and what ML architecture is used.
To study the transferability and also the robustness of our pretrained latent space, we project data available in the literature onto our model and study the corresponding structure dynamics correlation. We choose the  GlassBench dataset available in Ref.~\cite{Jung2025Roadmap}.  It provides 100 initial particle configurations and corresponding 
precomputed isoconfigurational displacement propensities at multiple
times for the three dimensional 80:20  Kob Andersen model ($\chi_{B}=0.2)$ at $T=0.44$. It also provides the SBO (Smooth bond order) descriptor and its coarse grained version, SBO CG obtained using unsupervised learning (PCA) for the A particles.
We use the coordinates of the 100 initial configurations from the GlassBench dataset in Ref.\cite{Jung2025Roadmap} and compute their SOAP vectors with $r_{cut}=1.4$, averaged over nearest neighbours. We then project the averaged SOAP output onto our pretrained models and use the latent parameter $\lambda_{1}$ and correlate it with the dynamics as given in Ref.\cite{Jung2025Roadmap}. Here we use both residual AE and Graph AE latent space prediction as they show the best performance. To quantify the structure dynamics correlation following Ref.\cite{Jung2025Roadmap}, we calculate the Pearson correlation coefficients, $C_P$, plotted in   Fig.~\ref{fig:glassbench_pearson} (a) at different times.
It is defined as
$C_P =
\frac{\langle (X-\langle X\rangle)
(\mu-\langle\mu\rangle)\rangle}
{\sqrt{\langle(X-\langle X\rangle)^2\rangle
\langle(\mu-\langle\mu\rangle)^2\rangle}}$,
where $X$ represents the structural order parameter
predicted by the model and $\mu$ denotes the
isoconfigurational propensity.
For comparison, the results of earlier studies by Coslovich {\it et al.} \cite{Coslovich2022JCP} and Berthier {\it et al} \cite{Jung2025Roadmap} are also presented, where SBO and SBO CG are used, respectively. We find that both the Residual AE and the Graph AE outperform SBO and SBO CG. 
This correlation, as shown here, can be obtained only when the SOAP vectors contain both radial and orientational information and are averaged over local neighbours \cite{Donkor2024JPCL,Lionello2025JCTC}. In Fig.~\ref{fig:glassbench_pearson} we plot the Pearson correlation of coarse grained order parameters and the dynamics at $\tau_{\alpha}$. We find that the prediction of all the order parameters further improves when they are coarse grained to their optimum value at $L=L_{max}$. Although at $t=\tau_{\alpha}$ and $L=0$ the Graph AE performs marginally better than the Residual AE, after coarse graining the results reverse.

Note that the projected data here are from an NVT simulation for N=4000, T=0.44 and our model is trained on NPT simulation studies for N=16000, from T=5.0-0.56, and $\chi_{B}=0.0-0.5$, showing the transferability of our trained model across simulation conditions (NPT vs. NVT), system size, composition and temperature range. 

One of the best performimg unsupervised learning models, the time lagged autoencoder (TAE), uses radial and orientational descriptors similar to SOAP. It shows that, even without any coarse graining, the structure dynamics correlation is good at long times \cite{QiuYethiraj2025PNAS}. Note that although TAE is unsupervised, it is trained on temporally paired configurations where the pairing was done over short time intervals. Thus, it uses some dynamical information. Our models, on the other hand, are trained solely on static structural snapshots. In our case, dynamics enters through the choice of optimum coarse graining length, $L_{\mathrm{max}}$, obtained by maximising the structure dynamics correlation. Our results therefore suggest that the TAE effectively performs a spatial smoothing comparable to our explicit coarse graining, with the length scale learned during training in one case and tuned afterwards in the other.

In the TAE study, it is also found that particles beyond the nearest neighbours contribute to the dynamics~\cite{QiuYethiraj2025PNAS}. Although the correlation of our model is best for a SOAP cutoff $r_{\mathrm{cut}}=1.4$, this is consistent with our observation that, at lower temperatures, the structure dynamics correlation grows appreciably only upon coarse graining over larger length scales, suggesting that neighbours beyond the first shell contribute to the long time dynamics. We further find that $L_{\mathrm{max}}$ grows as the temperature is lowered, consistent with a growing dynamic correlation length 
\cite{BerthierBiroli2011RMP, Karmakar2010PRL, TahKarmakar2020PRResearch, Paul2023PNAS, Donati1999PRL,FlennerSzamel2013JCP,Hallett2018NatCommun,Scalliet2022PRX}. It would be interesting to test whether the effective spatial range of the TAE features shows a similar growth as temperature decreases.

\section{Conclusions}
\label{sec:conclusions}

Our results show that a structural representation learned from locally averaged SOAP vectors, without any dynamical information, can represent the local order in different binary mixtures. When cooling trajectories are projected onto the latent space, the Residual AE places the crystallising compositions in the fcc/hcp and CsCl basins and keeps the frustrated compositions $\chi_B=0.2$--$0.4$ in the liquid region. This is the same classification obtained from BOO \cite{Steinhardt1983PRB,LechnerDellago2008JCP} and TCC \cite{Malins2013Faraday}, so the learned coordinate captures the crystal like order that competes with glass formation. The training data, although it does not contain pure crystalline structures, contains liquid to crystal trajectories, which alone are sufficient for the crystal basins to emerge. Most importantly, unlike BOO, the learned coordinate also resolves structural differences within the liquid. In the BOO space, the liquid trajectories of different compositions and temperatures nearly overlap, whereas in the Residual AE space they are well separated. We attribute this additional resolution mainly to the radial information in SOAP, which the BOO parameters lack. This additional sensitivity matters because the frustrated compositions, which never crystallise during cooling, are exactly the ones where BOO and TCC show little beyond a liquid like state, yet where the competition between crystal structures still shapes the liquid. The residual AE space also shows the crystallisation tendencies of different frustrated compositions, which are consistent with earlier observations \cite{Ujjwal_frust}.

Thus, the latent space not only identifies the crystal like order but can also identify finer structural changes in the liquid. This second sensitivity is what makes it relevant in predicting the dynamics of the supercooled liquid. The MLOP correlates with isoconfigurational mobility, whereas $Q_6$ does not. The correlation is highest when the SOAP vectors contain both radial and orientational information and are locally averaged. Among the architectures compared, the Residual AE is the simplest architecture, which gives the best result. We find that the Graph AE also gives comparable and, in some cases, marginally better results than the Residual AE model.

The comparison with the microscopic caging potential provides a physical picture to the learned latent order parameter. The predictions of the MLOP and of the caging potential are very similar: the two parameters are moderately correlated at the single particle level and strongly correlated once coarse grained. The species dependence of the structure dynamics correlation in the $\chi_B=0.2$ and $0.4$ mixtures is identically predicted by both the order parameters. Compared to the bare caging potential the bare MLOP shows a slightly higher correlation with the dynamics, which we attribute to the orientational information in SOAP that the microscopic theory does not use. It therefore appears that the MLOP behaves like a local caging potential that additionally contains orientational information. This is notable because no information about the interactions or the dynamics was provided to the model.

\begin{acknowledgments}

S.M.B. acknowledges funding from ANRF
and CSIR NCL Pune and thanks Giuseppe Foffi, 
and Biman Jana for scientific  discussions, and CSIR-4PI
for computational facilities.
S.K. thanks Mahesh Bhosale (University at Buffalo) and Saientan Bag
(Schrödinger Inc.) for useful discussions
on machine learning architectures, and DST-INSPIRE and CSIR-NCL
(Project No. MLP00262902) for fellowship support.

\end{acknowledgments}

\section*{Data availability}
The data and codes that support the findings of this study are
available from the corresponding author upon reasonable request.

\section{SUPPLEMENTARY INFORMATION}

\setcounter{figure}{0}
\renewcommand{\thefigure}{S\arabic{figure}}

\begin{figure}[t]
\centering
\includegraphics[width=\columnwidth]
{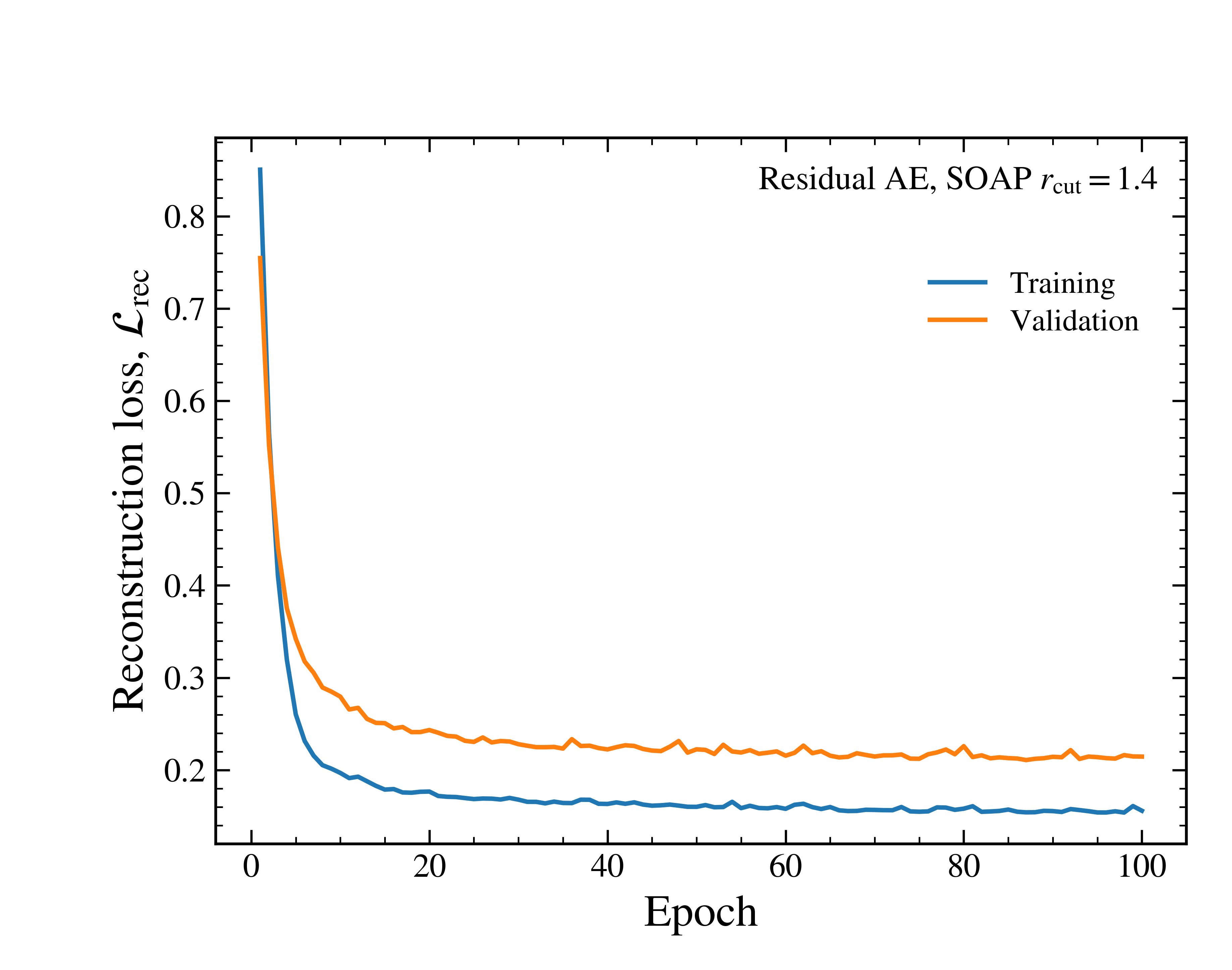}
\caption{
Training and validation reconstruction losses
for the Residual AE with
$r_{\rm cut}=1.4$.
The checkpoint corresponding to the
minimum validation loss is retained
for subsequent structural analysis.
}
\label{fig:ae_training_loss}
\end{figure}

\begin{figure}[t]
    \centering
    \includegraphics[width=\linewidth]
    {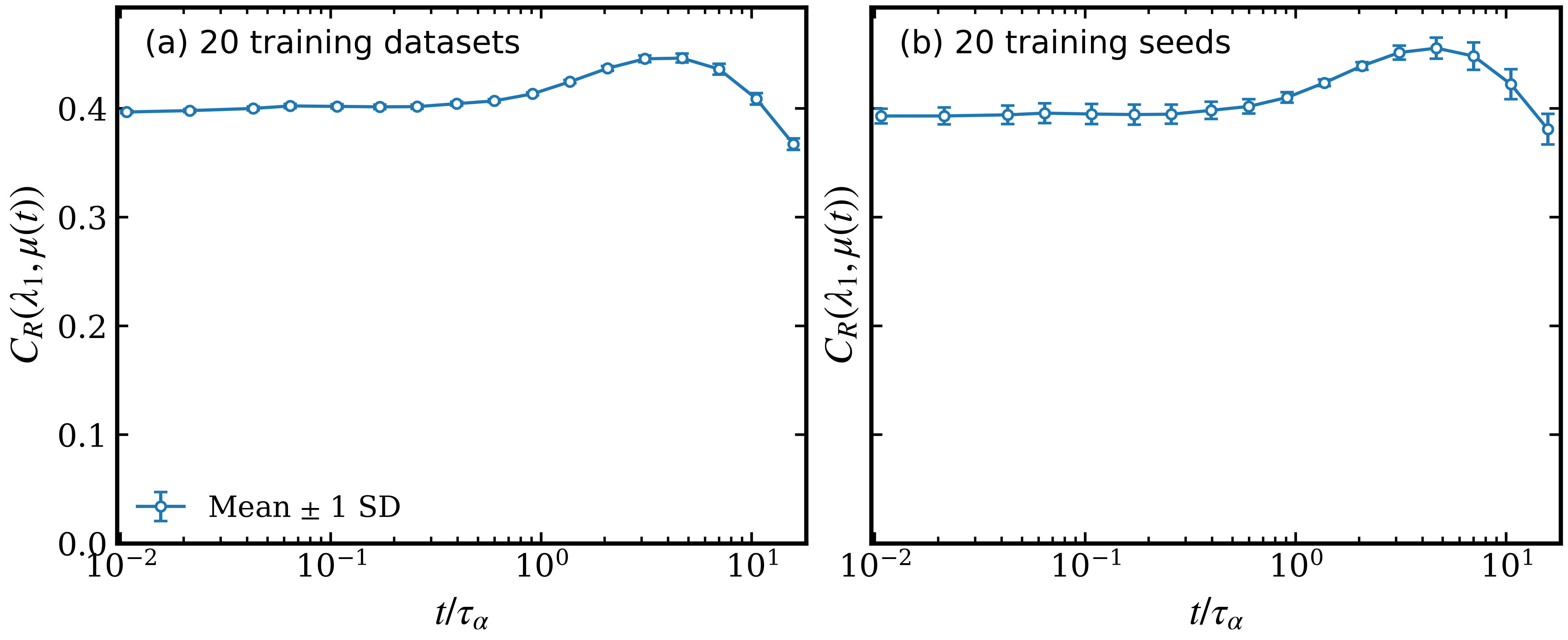}
\caption{
Sensitivity of the Residual AE structure dynamics
correlation to variations in the training dataset
and different seeds in the
$\chi_B=0.20$ mixture at $T=0.45$,
$\rho=1.20$, for A particles.
The absolute Spearman rank correlation
between the latent coordinate $\lambda_1$
and isoconfigurational mobility $\mu_i(t)$
is plotted as a function of $t/\tau_\alpha$.
Correlations are calculated separately for
each configuration, and their absolute values
are averaged over 12 independent configurations.
(a) Results from 20 training
datasets with the training seed fixed at 42.
(b) Results from 20 different training seeds using a fixed
training dataset. The two dimensional latent
spaces are aligned to that of reference seed 42
using orthogonal Procrustes alignment before
calculating the correlation with mobility.
The alignment uses only structural latent
coordinates and does not involve any dynamical
information. All 20 seeds are included.
Symbols and error bars represent the mean
and one sample standard deviation across
the 20 datasets in (a) and the 20 aligned
seeds in (b).
}
    \label{fig:ae_dataset_seed_variation}
\end{figure}

\begin{figure*}[t]
    \centering
    \includegraphics[width=0.90\textwidth]
    {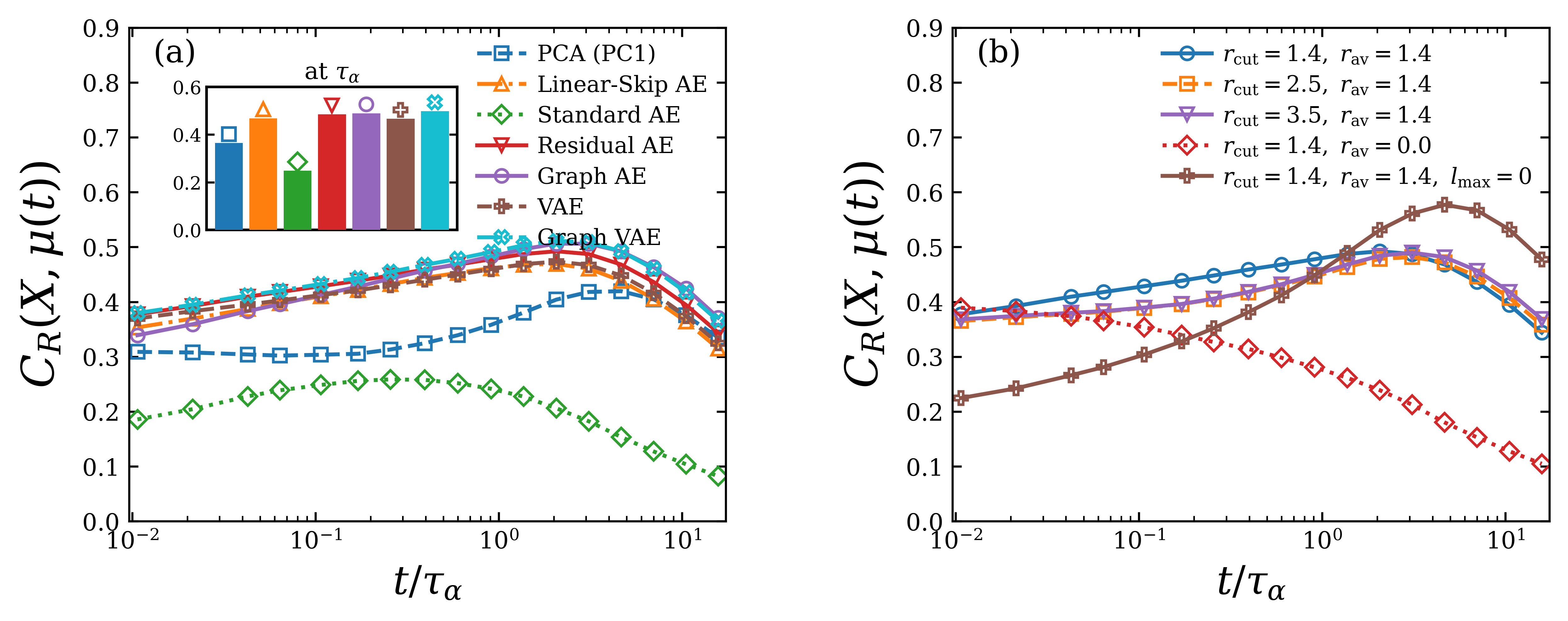}
\caption{Absolute Spearman rank correlation
$C_R(X,\mu(t))$ between structural descriptors
and isoconfigurational mobility $\mu_i(t)$
of B particles in the $\chi_B=0.20$ mixture
at $T=0.45$ and $\rho=1.20$, as a function
of $t/\tau_\alpha$.
(a) Comparison of PCA (PC1), Linear Skip AE,
Standard AE, Residual AE, Graph AE, VAE,
and Graph VAE (both with $\beta_{\max}=10^{-4}$).
(b) Comparison of Residual AE models trained
using different SOAP descriptor settings:
$r_{\mathrm{cut}}=1.4$, $2.5$, and $3.5$,
each with neighbour averaging over
$r_{\mathrm{av}}=1.4$;
$r_{\mathrm{cut}}=1.4$ without neighbour
averaging ($r_{\mathrm{av}}=0$);
and radial only SOAP descriptors
($l_{\max}=0$) with $r_{\mathrm{cut}}=1.4$
and $r_{\mathrm{av}}=1.4$.
Correlations are calculated separately
for each configuration, and their absolute
values are averaged over 12 independent
configurations.}
    \label{fig:B_type_structure_dynamics}
\end{figure*}

\begin{figure*}[t]
    \centering
    \includegraphics[width=\textwidth]
    {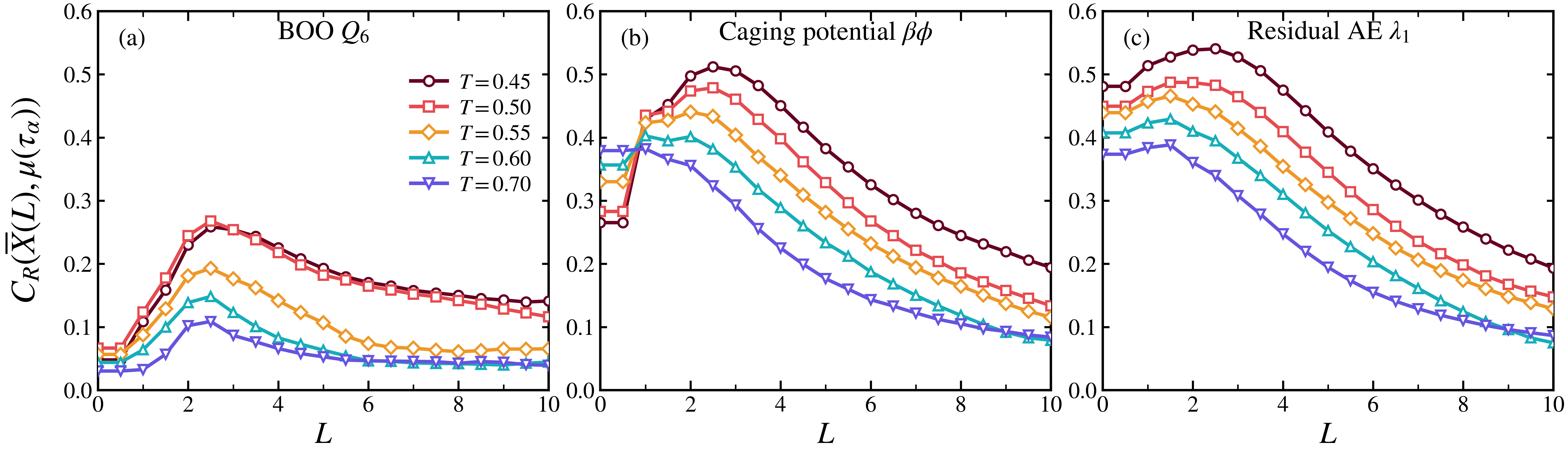}
\caption{Absolute Spearman rank correlation
$C_R(\overline{X}(L),\mu(\tau_\alpha))$
between the coarse grained structural order parameter and
isoconfigurational mobility $\mu$ at $t=\tau_\alpha$, as a function
of the coarse graining length $L$
(Eq.~(21) in the main text), for B particles in the
$\chi_B=0.20$ mixture at $\rho=1.2$.
The panels show (a)~$X=Q_6$, the bond orientational order
parameter, (b)~$X=\beta\phi$, the caging potential, and
(c)~$X=\lambda_1$, the Residual AE coordinate.
Colours and markers denote temperatures from $T=0.45$ to
$T=0.70$. The caging potential and the Residual AE coordinate
show maxima at a finite coarse graining length $L_{\max}$
and stronger correlations than $Q_6$.
Correlations are calculated separately for each configuration,
and their absolute values are averaged over 12 independent
configurations.}
    \label{fig:8020_B_coarsegrain}
\end{figure*}

\begin{figure*}[t]
\centering
\includegraphics[width=\textwidth]
{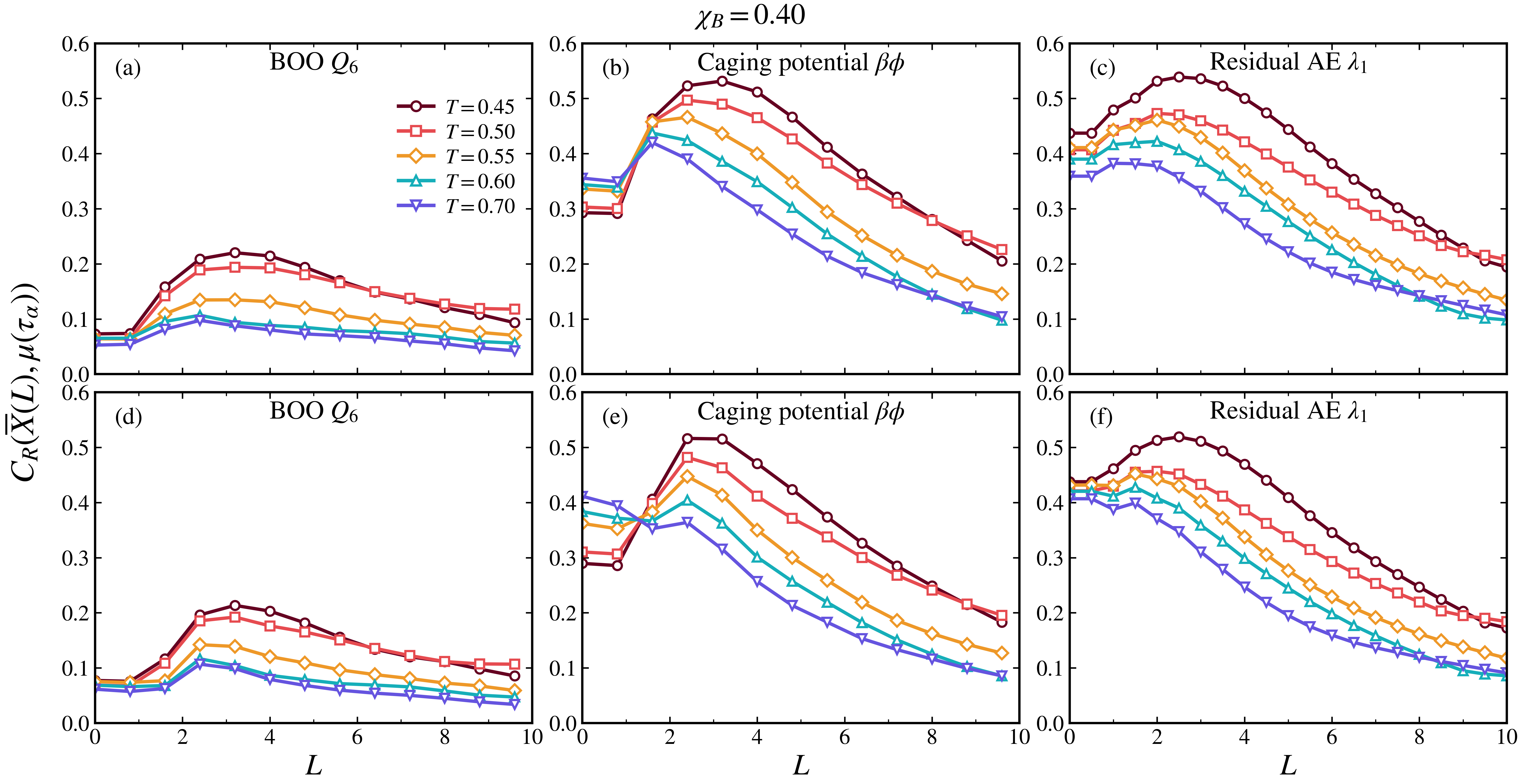}
\caption{Absolute Spearman rank correlation
$C_R(\overline{X}(L),\mu(\tau_\alpha))$
between the coarse grained structural order parameter and
isoconfigurational mobility $\mu$ at $t=\tau_\alpha$, as a function
of the coarse graining length $L$
(Eq.~(21) in the main text), for the
$\chi_B=0.40$ mixture at $\rho=1.41$.
The top row (a--c) corresponds to A type particles,
and the bottom row (d--f) corresponds to B type particles.
The columns show $X=Q_6$, the bond orientational order
parameter (a,d), $X=\beta\phi$, the caging potential (b,e),
and $X=\lambda_1$, the Residual AE coordinate (c,f).
Colours and markers denote temperatures from $T=0.45$ to
$T=0.70$. The caging potential and the Residual AE coordinate
show maxima at a finite coarse graining length $L_{\max}$
and stronger correlations than $Q_6$.
Correlations are calculated separately for each configuration,
and their absolute values are averaged over 12 independent
configurations.}
\label{fig:6040_coarsegrain}
\end{figure*}

\subsection*{S1. Residual autoencoder architecture and training details}
\label{app:ml_details}

We characterize the local environment of each particle
using the smooth overlap of atomic positions (SOAP)
descriptors, as described in Sec.~IV~C in the main text.
The descriptors are calculated with DScribe using
Gaussian type orbital radial basis functions,
periodic boundary conditions, and the parameters
$r_{cut}=1.4$, $n_{\max}=8$,
$\ell_{\max}=8$, and $\sigma_{\rm SOAP}=0.25$.
We treat A and B particles as a single species
in the SOAP calculation, yielding descriptors
of dimension $D=324$ that contain radial and orientational information.
Each SOAP descriptor is then averaged over
the central particle and its neighbours within
$r_{\rm av}=1.4$ denoted by $\bar{\mathbf{p}}$ (Eq.~(10) in the main text) to mininimized the local fluctuations.

The training dataset is constructed from NPT simulations at P = 10.19 of the Kob Andersen model at various compositions, $\chi_B = 0.0,\ 0.1,\, 0.2,\, 0.3,\, 0.4,\, 0.5$~\cite{Pedersen2018PRL}. A-rich compositions ($\chi_B \to 0,0.1$) crystallize into the fcc/hcp basin, equimolar compositions ($\chi_B = 0.5$) into the CsCl basin, and intermediate compositions are frustrated. For each of the six
compositions, we select one configuration at
each of 21 temperatures. All particles from
each configuration are included in the dataset.
The sampled descriptors are pooled across
temperatures and compositions to calculate the
mean $m_k$ and standard deviation $s_k$ of
each SOAP component $k$. The locally averaged SOAP
descriptor is standardized according to
$(\bar p_k^{(i)}-m_k)/s_k$, where $i$ indexes the particle. 
The same standardization parameters are used
for training, validation, and subsequent
evaluation. Note that within each composition, the
source trajectories are shuffled using seed 42,
and approximately 80\% are assigned to training
and 20\% to validation.

The Residual AE architecture is described
in Sec.~IV~D in the main text. Its encoder maps the
324-dimensional input through layers of widths
$324\rightarrow256\rightarrow256
\rightarrow128\rightarrow2$.
The first three transformations are residual
blocks, each containing a linear transformation,
layer normalization, SiLU activation, and dropout
with probability 0.05. The shortcut mapping
$\mathbf{S}^{(\ell)}$, defined in
Eq.~(13) in the main text, is the identity
when the input and output dimensions match
and a bias-free linear projection when
they differ. The final linear transformation
produces the two-dimensional latent
representation
$\mathbf z_i=(\lambda_{1,i},\lambda_{2,i})$. The decoder maps these coordinates back to
the original descriptor space through layers
of widths
$2\rightarrow128\rightarrow256
\rightarrow324$, with layer normalization
and SiLU activation in the two hidden layers
and a linear output layer.

The training target for each particle is
its own standardized, locally averaged
SOAP descriptor $\bar{\mathbf p}^{(i)}$.
The network parameters are optimized by
minimizing the mean squared reconstruction
error $\mathcal{L}_{\rm AE}$ defined in
Eq.~(14) in the main text, where the decoder
output is $\hat{\bar{\mathbf p}}^{(i)}$.
We update the encoder and decoder parameters
after each configuration using the AdamW
optimizer, with a learning rate of $10^{-4}$,
weight decay of $10^{-5}$, and a maximum
gradient norm of $1.0$. A single set of
network parameters is shared across
all temperatures and compositions.

Training is performed for 100 epochs, with
the training configurations reshuffled at
each epoch. At the end of each epoch,
we evaluate the reconstruction loss on
the validation configurations and retain
the network parameters corresponding to
the lowest validation loss.
Figure~\ref{fig:ae_training_loss} shows
the training and validation reconstruction
losses, each averaged over particles.
The entire training procedure uses only
structural descriptors; no information
about particle mobility is used.

For the structure dynamics analysis, the trained encoder is used to project independent NVT configurations at $\chi_B = 0.2$ and $0.4$ (with $\rho = 1.2$ and $1.41$, respectively) onto the latent space. At each temperature, the results are averaged over 12 independent initial configurations, each of which is used to generate an isoconfigurational ensemble of trajectories. The same SOAP parameters, local averaging, and standardization procedures used for training are applied.
Dropout is disabled during the projection.
The encoder provides the latent coordinates
$(\lambda_{1,i},\lambda_{2,i})$ for each
particle. We use the first coordinate,
$\lambda_1$, as the machine learned
structural order parameter and evaluate
its correlation with isoconfigurational
mobility $\mu_i(t)$ separately for A
and B particles.

Finally, we investigate the sensitivity
of the structure dynamics correlation
to variations in the training dataset
and various seeds
(Fig.~\ref{fig:ae_dataset_seed_variation}).
The analysis is performed for A particles
in the $\chi_B=0.20$ mixture at $T=0.45$
and without additional spatial
coarse graining ($L=0$).

At fixed training seed 42, the absolute
Spearman rank correlation between
$\lambda_1$ and $\mu_i(t)$ varies little
across 20 training
datasets.

Changing the training seed leads to considerable variation in the correlation, even when the same training dataset is used. This variation can arise because independently trained autoencoders may produce differently oriented two dimensional latent spaces. To account for this, we align the latent spaces obtained from 20 different training seeds to that of a reference seed (seed 42), in which $\lambda_1$ follows the temperature evolution and $\lambda_2$ follows the transition toward the crystalline basins, using orthogonal Procrustes alignment~\cite{schonemann1966,gower2004procrustes}.
A single transformation is determined for
each seed using the latent representations
of the same particles across all 12
independent configurations. 
For each training seed, we first calculate the
mean of the two dimensional latent coordinates
and subtract it from the corresponding latent
representations, thereby centering the latent
space at the origin. We then apply an orthogonal
Procrustes transformation to rotate or reflect
the centered latent coordinates so that they
best match those obtained from the reference
model trained with seed 42. Finally, we add
the mean latent coordinates of the reference
model to the transformed coordinates, ensuring
that the aligned latent spaces have the same
center and a consistent orientation.
The alignment uses only the structural
latent coordinates and does not involve
any dynamical information.

Following alignment, the structure dynamics
correlation is calculated using the aligned
latent coordinate $\lambda_1$ which is representing the temperature evolution in Residual AE latent space. 
The results show very small variation
in the structure dynamics correlation
across the 20 training seeds (Fig. S2 (b)).

\subsection*{S2. Structure dynamics correlations of B particles in the $\chi_B=0.2$ mixture}
\label{app:8020B_corr}

In Figs.~5 and~7 of the main text we analyze the A particles of the
$\chi_B=0.20$ mixture. Here we repeat the analysis for B particles to test
whether the conclusions depend on the particle species.

Figure~\ref{fig:B_type_structure_dynamics} shows the absolute Spearman rank
correlation between structural descriptors and isoconfigurational mobility
as a function of $t/\tau_\alpha$ for B particles at $T=0.45$, without
additional coarse graining of the trained structural descriptors ($L=0$).
Panel~(a) compares PCA (PC1), Linear-Skip AE, Standard AE, Residual AE,
Graph AE, VAE, and Graph VAE. Panel~(b) examines the effect of SOAP settings
on the Residual AE predictions: cutoffs $r_{\rm cut}=1.4$, $2.5$, and $3.5$
with neighbour averaging over $r_{\rm av}=1.4$, descriptors without neighbour
averaging ($r_{\rm av}=0$), and radial only descriptors ($\ell_{\max}=0$).
These results show that both the residual term in the architecture and the
local averaging of SOAP descriptors are important.

Figure~\ref{fig:8020_B_coarsegrain} shows the absolute Spearman rank correlation between structural descriptors and isoconfigurational mobility at  $t=\tau_\alpha$ for B particles
as a function of the coarse-graining length $L$ for temperatures from
$T=0.45$ to $T=0.70$. The three panels correspond to the bond orientational
order parameter $Q_6$, the caging potential $\beta\phi$, and the Residual AE
coordinate $\lambda_1$. As observed for A particles, $Q_6$ also correlates only weakly with mobility for B particles, whereas $\beta\phi$ and $\lambda_1$ exhibit stronger correlations.
For these two descriptors, the correlation rises with $L$, peaks at a finite
$L_{\max}$, and then decreases, and it strengthens upon cooling. At $L=0$,
the Residual AE performs better than the other descriptors.

\subsection*{S3. Structure dynamics correlations in the $\chi_B=0.4$ mixture}
\label{app:6040_corr}

We also examine the structure dynamics correlations of the
$\chi_B=0.4$ mixture, in which A and B particles have similar
environments. Figure~\ref{fig:6040_coarsegrain} shows the correlation
between the structural descriptors and the isoconfigurational mobility at $t=\tau_\alpha$ as
a function of $L$, separately for A and B particles at different
temperatures.

For both species, $Q_6$ is only weakly correlated with the isoconfigurational  mobility,
whereas the caging potential and the Residual AE coordinate $\lambda_1$ show much
stronger correlations. For both of these descriptors the correlation
increases with $L$ up to a maximum at a finite $L$ and then decreases. It
also becomes stronger on cooling, with the largest values at $T=0.45$.
The Residual AE coordinate gives slightly higher peak correlations than the
caging potential for both species, and A and B particles behave
similarly.

These results show that structural information beyond the first neighbour shell is important for predicting the future mobility of particles. However, the non monotonic dependence on $L$ shows that averaging over too large a region washes out the local structural information that governs mobility, which weakens the structure dynamics correlation.

\subsection*{S4. Other machine learning techniques used in the main paper}
\label{app:other_ml_models}

In addition to the Standard AE, Linear Skip AE,
and Residual AE described in
Sec.~IV~D in the main text, we consider a graph
autoencoder (Graph AE), a variational
autoencoder (VAE), and a graph variational
autoencoder (Graph VAE). All these models
are trained to reconstruct the standardized,
locally averaged SOAP descriptors
$\bar{\mathbf p}^{(i)}\in\mathbb{R}^{324}$
without using any dynamical information.
Graph based neural networks have been
used to study glassy dynamics because
they incorporate information from
neighbouring particles through message
passing
\cite{Bapst2020NatPhys,Shiba2023JCP,
Pezzicoli2024SciPost}.
The VAE provides a probabilistic latent
representation regularized towards a
specified prior distribution, allowing us
to examine the influence of latent-space
regularization on the learned structural
order parameter
\cite{KingmaWelling2014,
MaityChakrabarty2025JCTC}.

\subsubsection{Graph autoencoder (Graph AE)}
\label{app:graph_ae}

\begin{figure}[t]
    \centering
    \includegraphics[width=\columnwidth]{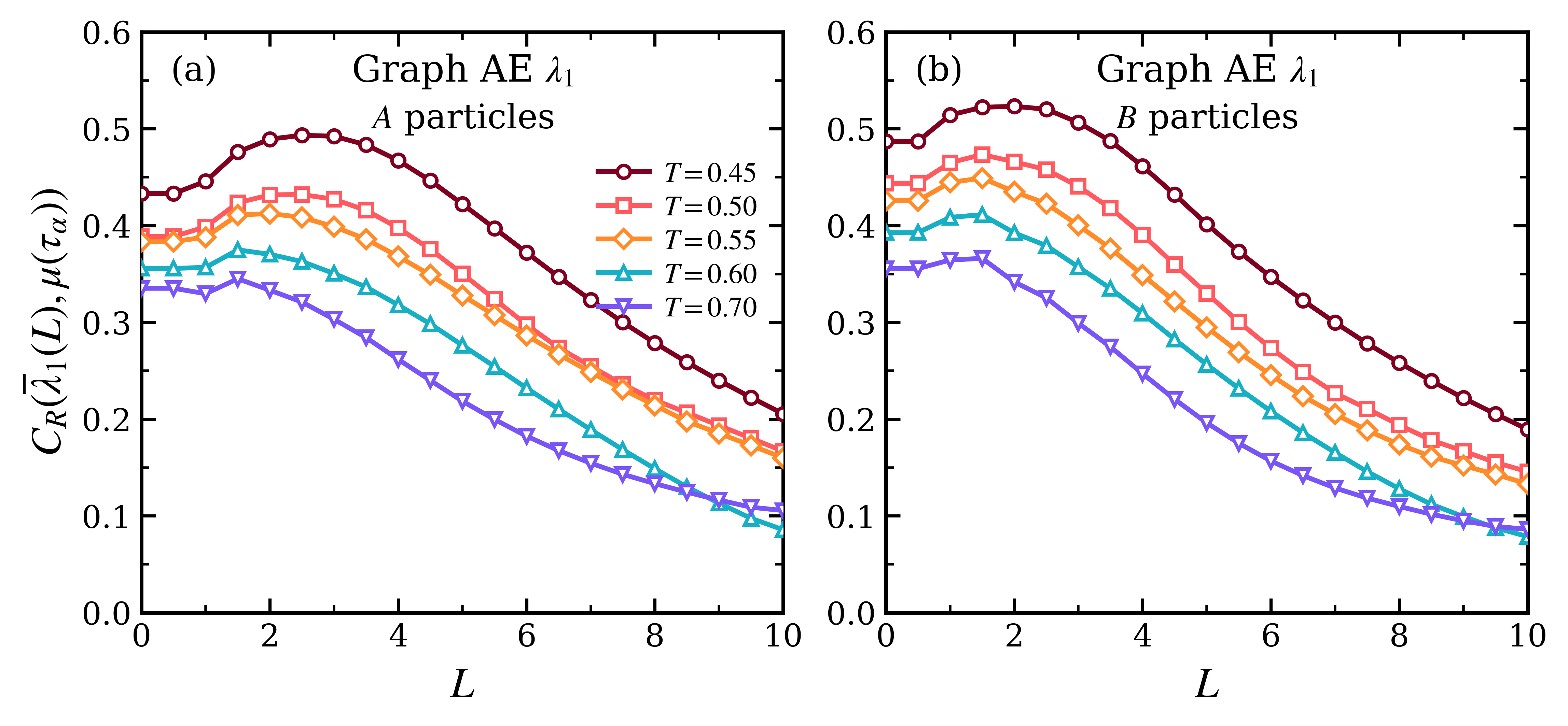}
\caption{Absolute Spearman rank correlation
$C_R(\overline{\lambda}_1(L),\mu(\tau_\alpha))$ between the
coarse-grained Graph AE coordinate and isoconfigurational
mobility $\mu$ at $t=\tau_\alpha$, as a function of the
coarse-graining length $L$, for (a) A particles and
(b) B particles in the $\chi_B=0.20$ mixture at $\rho=1.2$.
Colours and markers denote temperatures from $T=0.45$ to
$T=0.70$. The Graph AE coordinate shows maxima at a finite
coarse-graining length $L_{\max}$ for both particle types.
Correlations are calculated separately for each configuration,
and their absolute values are averaged over 12 independent
configurations.}
    \label{fig:graph_ae_coarse_graining}
\end{figure}

The graph autoencoder, referred to as Graph AE
in the main text, uses a Graph Attention Network
(GAT) to incorporate information from neighbouring
particle environments through graph message
passing with learned attention weights
\cite{Velickovic2018GAT}.

For each configuration, particles form
the nodes of a periodic radius graph.
Each node $i$ carries the standardized,
locally averaged SOAP descriptor
$\bar{\mathbf p}^{(i)}$, defined in
Sec.~IV~C in the main text. Two distinct particles
are connected when their minimum image
separation satisfies $r_{ij}<r_g$,
where $r_g$ is the graph cutoff.
Each connected pair is represented by
directed edges in both directions, and
a self edge is included for every particle.

Starting from
$\mathbf{h}_i^{(0)}=\bar{\mathbf p}^{(i)}$,
the encoder first processes the SOAP descriptors
through two residual MLP blocks, followed by
a graph attention layer with four attention heads.
The aggregated message is expressed as
\begin{equation}
\mathbf{m}_i^{(\ell)}
=
\mathop{\Vert}_{a=1}^{4}
\left(
\sum_{j\in\mathcal{N}_g(i)\cup\{i\}}
\alpha_{ij,a}^{(\ell)}
\mathbf{W}_a^{(\ell)}
\mathbf{h}_j^{(\ell)}
\right),
\label{eq:gat_message}
\end{equation}
where $\Vert$ denotes concatenation,
$\mathcal{N}_g(i)$ is the set of graph
neighbours of particle $i$, and $a$
indexes the four attention heads.
Here, $\mathbf{W}_a^{(\ell)}$ is a trainable
feature transformation and
$\alpha_{ij,a}^{(\ell)}$ is the attention
coefficient assigned to particle $j$
by attention head $a$. The attention
coefficients are normalized over the
neighbouring particles and the central
particle for each attention head,
determining their relative contributions
to the aggregated message.

The resulting features are passed through
a third residual MLP block and a final
linear transformation to obtain the
two dimensional latent representation
$\mathbf{z}_i=(\lambda_{1,i},\lambda_{2,i})$.
A node wise decoder reconstructs
$\hat{\bar{\mathbf p}}^{(i)}$
from $\mathbf{z}_i$ without further
graph message passing.

The Graph AE is trained by minimizing
the mean squared reconstruction error
$\mathcal{L}_{\rm AE}$ defined in
Eq.~(14) in the main text. As in the
other autoencoder architectures,
the loss is evaluated against the
input averaged SOAP descriptors,
and no dynamical information is
used during training.

Figure~\ref{fig:graph_ae_coarse_graining} shows the
absolute Spearman rank correlation between the
coarse-grained Graph AE coordinate
$\overline{\lambda}_1(L)$ and the isoconfigurational
mobility at $t=\tau_\alpha$ for A and B particles
at different temperatures. For both particle types,
the correlation initially increases with the
coarse graining length $L$, reaches a maximum at
a finite $L_{\max}$, and subsequently decreases.
The correlations generally become stronger with
decreasing temperature, highlighting the importance
of intermediate range structural correlations in
capturing dynamical heterogeneity.

\begin{figure}[t]
    \centering
    \includegraphics[width=\columnwidth]{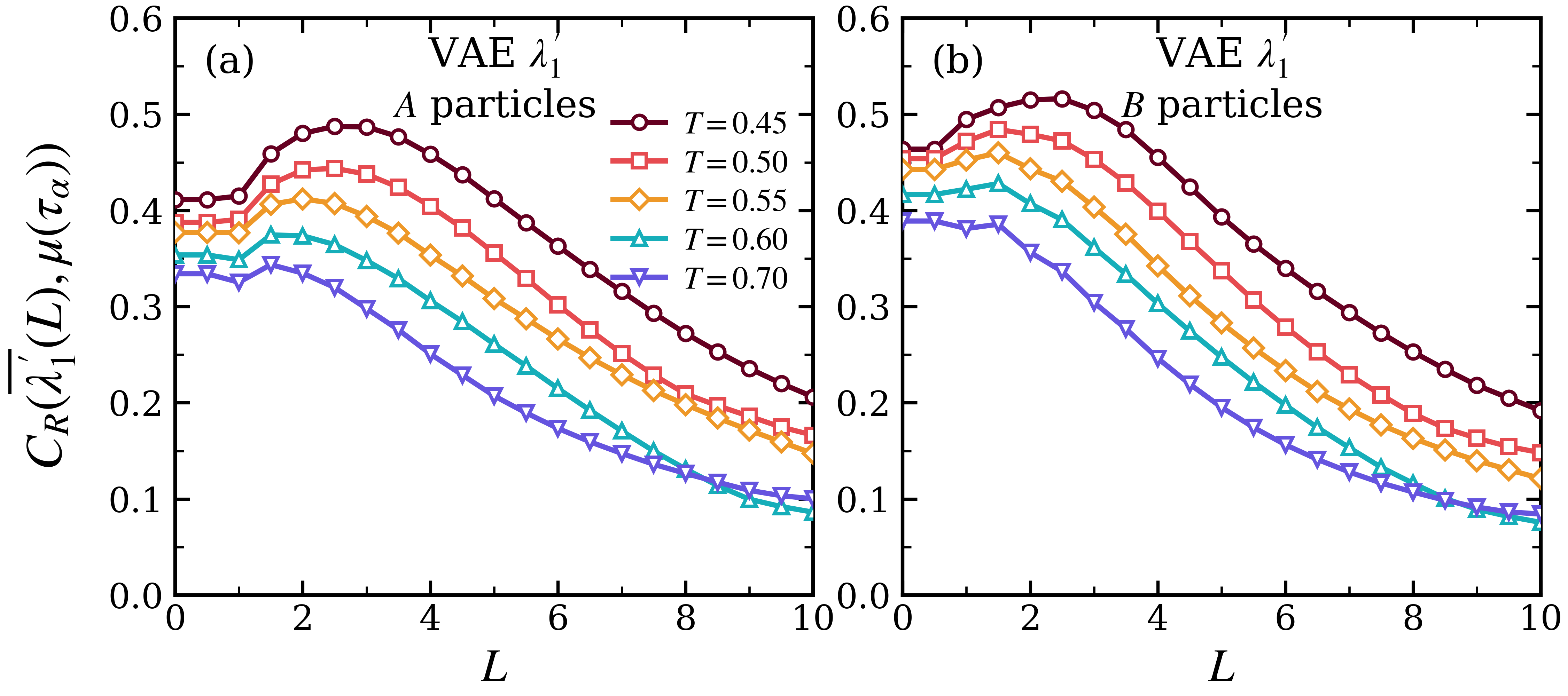}
    \caption{Absolute Spearman rank correlation
    $C_R(\overline{\lambda_1}(L),\mu(\tau_\alpha))$
    between the coarse grained VAE coordinate and isoconfigurational
    mobility $\mu$ at $t=\tau_\alpha$, as a function of the coarse graining
    length $L$, for (a) A particles and
    (b) B particles in the $\chi_B=0.20$ mixture at $\rho=1.2$.
    The VAE is trained using SOAP descriptors with $r_{\mathrm{cut}}=2.5$.
    Colours and markers denote temperatures from $T=0.45$ to $T=0.70$.
    The VAE coordinate shows maxima at a finite coarse graining length
    $L_{\max}$ for both particle types.
    Correlations are calculated separately for each configuration, and
    their absolute values are averaged over 12 independent configurations.}
    \label{fig:vae_coarse_graining}
\end{figure}

\subsubsection{Variational autoencoder (VAE)}
\label{app:vae}

Unlike the deterministic autoencoders
described in Sec.~IV~D in the main text,
the VAE maps each locally averaged and standardized  SOAP
descriptor to a probability distribution
in the two dimensional latent space.
The encoder predicts the mean
$\boldsymbol{\mu}_i$ and variance
$\boldsymbol{\sigma}_i^2$ of a
diagonal Gaussian distribution,
\begin{equation}
q_{\theta}
\left(
\mathbf{z}_i\mid
\bar{\mathbf p}^{(i)}
\right)
=
\mathcal{N}
\left(
\boldsymbol{\mu}_i,
\operatorname{diag}
(\boldsymbol{\sigma}_i^2)
\right),
\label{eq:vae_posterior}
\end{equation}
where $\theta$ denotes the encoder
parameters.

During training, a latent vector is
sampled using the reparameterization
trick,
\begin{equation}
\mathbf{z}_i
=
\boldsymbol{\mu}_i
+
\boldsymbol{\sigma}_i
\odot\boldsymbol{\epsilon}_i,
\qquad
\boldsymbol{\epsilon}_i
\sim
\mathcal{N}(\mathbf{0},\mathbf{I}),
\label{eq:vae_sampling}
\end{equation}
where $\odot$ denotes element-wise
multiplication. This formulation
allows gradients to propagate through
the sampling operation.

The decoder reconstructs the averaged
SOAP descriptor from the sampled
latent vector $\mathbf{z}_i$.
Training minimizes the reconstruction
error together with a
Kullback--Leibler (KL) divergence term,
\begin{equation}
\mathcal{L}_{\rm VAE}
=
\mathcal{L}_{\rm AE}
+
\beta_{\rm KL}
\left\langle
D_{\rm KL}
\left[
q_{\theta}
\left(
\mathbf{z}_i\mid
\bar{\mathbf p}^{(i)}
\right)
\,\Vert\,
\mathcal{N}(\mathbf{0},\mathbf{I})
\right]
\right\rangle_i,
\label{eq:vae_loss}
\end{equation}
where $\mathcal{L}_{\rm AE}$ is
the reconstruction loss defined in
Eq.~(14) in the main text, and
$\beta_{\rm KL}$ controls the
strength of latent-space
regularization. The KL term
penalizes deviations of the
posterior distributions from
the standard Gaussian prior.

The finite $L$ maxima in Fig.~\ref{fig:vae_coarse_graining}
suggest that incorporating intermediate range structural
information enhances the ability of the VAE coordinate
to capture dynamical heterogeneity, particularly at
lower temperatures.

\subsubsection{Graph variational autoencoder (Graph VAE)}
\label{app:graph_vae}

\begin{figure}[t]
    \centering
    \includegraphics[width=\columnwidth]{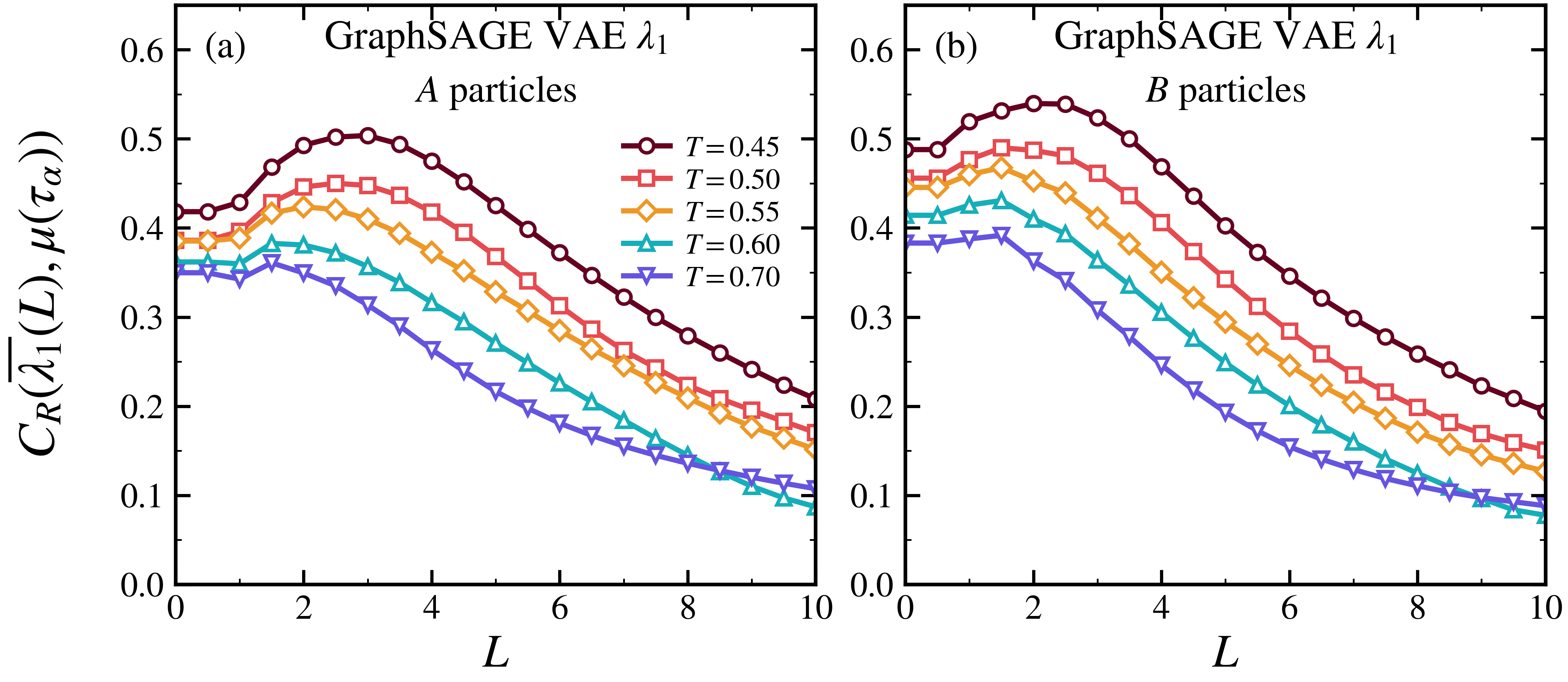}
    \caption{Absolute Spearman rank correlation
    $C_R(\overline{\lambda_1}(L),\mu(\tau_\alpha))$
    between the coarse-grained Graph VAE coordinate and
    isoconfigurational mobility $\mu$ at $t=\tau_\alpha$, as a function
    of the coarse-graining length $L$, for (a) A particles and
    (b) B particles in the $\chi_B=0.20$ mixture at $\rho=1.2$.
    The Graph VAE is trained using SOAP descriptors with
    $r_{\mathrm{cut}}=1.4$ and a graph cutoff of $2.5$.
    Colours and markers denote temperatures from $T=0.45$ to $T=0.70$.
    The correlation reaches a maximum at a finite coarse-graining length
    $L_{\max}$ for both particle types.
    Correlations are calculated separately for each configuration, and
    their absolute values are averaged over 12 independent configurations.}
    \label{fig:graphsage_vae_coarse_graining}
\end{figure}

The Graph VAE combines a GraphSAGE
(Graph Sample and Aggregate) encoder
\cite{Hamilton2017GraphSAGE} with the variational
formulation described in Sec. S4.2.
GraphSAGE generates node representations by
aggregating the features of neighbouring nodes
and combining them with those of the central node.
Here, the graph encoder incorporates information
from neighbouring SOAP environments through mean
aggregation and predicts the posterior mean
$\boldsymbol{\mu}_i$ and variance
$\boldsymbol{\sigma}_i^2$ for each particle.
The latent distribution is therefore conditioned
on the graph based representation of the local
structural environment.

During training, latent vectors are sampled using
Eq.~\eqref{eq:vae_sampling}.
A node-wise decoder reconstructs
the input averaged SOAP descriptors
without further graph message passing.
The model is optimized using the
reconstruction and KL divergence
terms defined in
Eq.~\eqref{eq:vae_loss},
with the posterior distribution
provided by the graph encoder.

The temperature dependence shown in
Fig.~\ref{fig:graphsage_vae_coarse_graining}
demonstrates the ability of the Graph VAE latent
coordinate $\lambda_1$ to capture dynamical heterogeneity
across the supercooled regime.

However, incorporating
graph based message passing does not lead to any
substantial improvement in the structure dynamics
correlation compared to the Residual AE.

%

\end{document}